\documentclass[amsmath,amssymb,aps,reprint,pra]{revtex4-1}

\usepackage{graphicx}
\usepackage{subfigure}     
\usepackage{braket}
\makeindex
\usepackage{rotating}

\usepackage{color}

\begin{document}
	
	\title[CQED and chiral quantum optics]{Cavity quantum electrodynamics and chiral quantum optics}\label{ra_ch1}
	
\author{Michael Scheucher$^1$}\email{michael.scheucher@tuwien.ac.at}
\author{J\"urgen Volz$^{2}$}\email{juergen.volz@hu-berlin.de}
\author{Arno Rauschenbeutel$^{2}$}\email{arno.rauschenbeutel@hu-berlin.de}
\affiliation{$^1$Vienna Center for Quantum Science and Technology, TU Wien-Atominstitut, Stadionallee 2, 1020 Vienna, Austria}
\affiliation{$^2$ Department of Physics, Humboldt-Universit\"at zu Berlin, Newtonstr. 15, 12489 Berlin, Germany.}

	\begin{abstract}
		
		Cavity quantum electrodynamics (CQED) investigates the interaction between light confined in a resonator and particles, such as atoms.
		In recent years, CQED experiments have reached the optical domain resulting in many interesting applications in the realm of quantum information processing.
		For many of these application it is necessary to overcome limitations imposed by photon loss. In this context whispering-gallery mode (WGM) resonators have obtained significant interest.  Besides their small mode volume and their ultra high quality, they also exhibit favorable polarization properties that give rise to chiral light--matter interaction. In this chapter, we will discuss the origin and the consequences of these chiral features and we review recent achievements in this area.
		
	\end{abstract}

	\maketitle
	
	\tableofcontents


	\section{Introduction}\label{ra_sec2}

	Whispering-gallery waves were first described by Lord Rayleigh in 1878\cite{Rayleigh:1910aa}. He was able to show that the ghostly whispers that had been heard in the gallery of St. Paul's Cathedral are sound waves that travel along the curved wall being reflected multiple times, thus, allowing to hear a whisper of another person all along the gallery despite the large distance to the speaker. When the waves interfere after a round trip, discrete modes can form. Sonic whispering-gallery waves can be found in extremely large objects, such as the earth or even the sun, as well as in very small objects, such as single atomic nuclei.	
	As whispering-gallery waves are in principle just waves being continuously reflected from a curved concave surface, they are not restricted to sound.
	The most important examples might be optical whispering-gallery modes (WGMs) that owe their existence to total internal reflection, which occurs when light tries to exit from a dielectric at a shallow angle. While optical WGMs have much in common with with their acoustic counterpart, there are some subtle difference. 
	While sound waves in air are longitudinal waves, that is they exclusively oscillate in the direction of propagation, light waves are transverse waves. As a consequence, optical WGMs can exhibit complex polarization properties. For example, the polarization can get linked to the propagation direction and thus be different for light circulating in one or the other direction around the circumference. This effect lies at the core of this chapter, where we will discuss the origin and the consequences of direction-dependent polarization on the coupling between light and matter.
	
	\section{Spin--momentum locking in WGM resonators}\label{ra_sec1}
	
	In the standard approximation of paraxial optics one makes the assumption that the light beams only vary slowly in the directions transverse to the beam's propagation direction. Consequently, the electromagnetic field of these beams oscillates solely in a plane perpendicular to the direction of propagation, forming a purely transverse wave.  However, this approximation fails when beams are strongly focused or confined in transverse direction. In this case, the fields can also exhibit longitudinal polarization components that oscillate in the direction of propagation. Moreover, the longitudinal field components are in quadrature (90$^\circ$ out of phase) with respect to the transverse component. As a consequence, for a quasilinearly polarized beam, the local electric field vector rotates around an axis that is perpendicular to its propagation direction thereby forming a locally elliptically polarized field. Even though these effects are well known, are being applied in microwave engineering \cite{Pozar1998} and have already been describe in 1959 for optical waves \cite{Richards1959}, they recently received a great deal of attention \cite{Spin15}.
	These phenomena, which are now often referred to as transverse spin angular momentum (SAM) of light, were intensively studied theoretically \cite{Berry2009, Barnett2002, Cameron2012, Bliokh2014, Bliokh2014a,Young2014}. Furthermore, recent experimental advances made it possible to directly examine and probe the local properties of light \cite{Marrucci2006,Banzer2013,Bauer2014aa,Bauer2015}. One important consequence of transverse SAM is spin--momentum locking (SML), which links the local polarization to the propagation direction of the light \cite{Bliokh2014a,Aiello2015}. Recently, this has been used for realizing directional channeling of light, which was demonstrated across various optical systems, including surface plasmon polaritons \cite{Mueller2013,Lin2013,Shitrit2013,OConnor2014}, tightly focused Gaussian beams \cite{Neugebauer2014}, optical nanofibers\cite{Petersen2014,Mitsch2014} or waveguides \cite{Coles2016}, photonic crystal waveguides \cite{Sollner2015,LeFeber2015} and even WGM resonators\cite{Junge2013,Rosenblum2016}. However, the far-reaching consequences these local field properties have on the interaction of light and matter were often not taken into account.  \\ 
	In the following section, we will give a short introduction on transverse SAM  and SML of light, and discuss their origin by introducing a intuitive explanation, that covers the essence of this phenomena. Then, we will discuss the emergence of transverse SAM in WGMs.
	
	
	\subsection{From field gradients to longitudinal fields}
	\label{sec:gradient}
	Any electromagnetic wave is fully described by Maxwell's equations. In particular, the electric field $\boldsymbol{E}$ has to fulfill Gauss's law which in an isotropic, charge- and current-free dielectric media is given by
	\begin{equation}
	\text{div}\, \boldsymbol{E}=0\;.
	\end{equation} 
	As an example, we now consider a monochromatic beam with frequency $\omega$ propagating through vacuum in positive $y$-direction. Its electric field is given by ${\boldsymbol{E}}={\boldsymbol{\mathcal{E}}} \exp{i(\omega t- |\boldsymbol{k}| y)}+c.c.$, where ${\boldsymbol{\mathcal{E}}}=(\mathcal{E}_{x},\mathcal{E}_{y},\mathcal{E}_{z})$ is the complex valued vector amplitude of the field and ${\boldsymbol{k}}$ its wave vector. In addition, we assume that $\boldsymbol{\mathcal{E}}$ varies only slowly along the $y$-direction, thus 
	\begin{equation}
	\partial \boldsymbol{E}/\partial y \approx- i|\boldsymbol{k}|\boldsymbol{\mathcal{E}} \exp{i(\omega t- |\boldsymbol{k}| y)}+c.c.\;.
	\end{equation}
	From Gau\ss's law we then obtain the longitudinal field component 
	\begin{equation}
	\mathcal{E}_{y}\approx-i\frac{\lambda}{2\pi}\left(\frac{\partial \mathcal{E}_{x}}{\partial x}+\frac{\partial \mathcal{E}_{z}}{\partial z}\right).
	\label{eqn2}
	\end{equation} 
	Equation (\ref{eqn2}) shows that a strong longitudinal field component will occur as soon as the amplitude of the transverse field component varies on the length scale of $|\boldsymbol{ k}|^{-1}\!=\!\lambda/2\pi$.
	The imaginary unit $i$ in Eq.~(\ref{eqn2}) imposes that the longitudinal field component is in quadrature to the transverse field components, i.e. has a phase retardation of $\pi/2$. For a quasilinearly polarized field, the light is thus elliptically polarized in the plane that contains the propagation direction. When the longitudinal and the transverse field components have the same amplitude, i.e $ |\mathcal{E}_{y}|\!=\!\sqrt{|\mathcal{E}_{x}|^2+|\mathcal{E}_{z}|^2}$, the local polarization is perfectly circularly polarized\\
	There is a large variety of situations that feature strong gradients and thus show significant longitudinal polarization components. This includes tightly focused Gaussian beams, two-wave interference and most importantly evanescent fields. In addition, it is also observed in near-fields of plasmonic systems or photonic crystal wave-guides.

	\subsection{Transverse spin angular momentum of light} 
	
	\begin{figure}
		\centering
		\includegraphics[width=0.8\columnwidth]{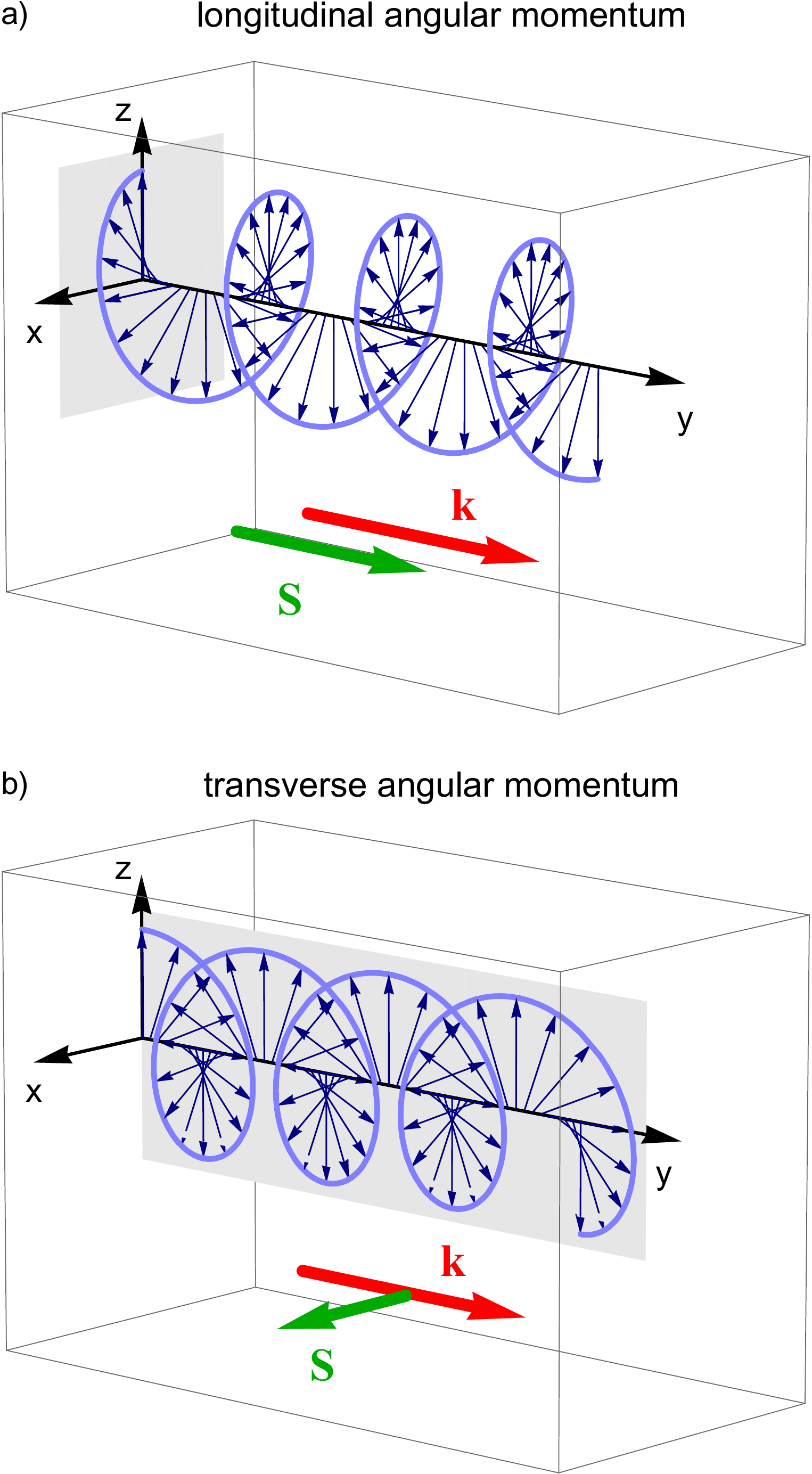}
		\caption{In order to illustrate the difference between transverse and longitudinal angular momentum of light we schematically show the electric field vectors (blue) for a given time along the wave's propagation direction ($y$-direction) for the two cases. a) For the case of longitudinal spin angular momentum, the field rotates about an axis parallel to the direction of propagation. This situation occurs, e.g. for a circularly polarized plane wave. b) For the case of transverse spin angular momentum, the field rotates in a plane spanned by the propagation direction and a transverse vector. In both diagrams the gray plane indicates the plane in which the field rotates. The red arrow represents the wave vector $\boldsymbol{k}$, and thus the propagation direction, while the green arrow indicates the local spin of the electric field $\boldsymbol{S}$, which is either a) parallel or b) perpendicular to $\boldsymbol{k}$. }
		\label{fig:trans_vs_long_angular_momentum}
	\end{figure}
	
	Already at the beginning of the 20th century, Pointing suggested to assign angular momentum to circularly polarized light \cite{Poynting1909}. Today, we distinguish two kinds of optical angular momenta: Orbital angular momentum and spin angular momentum (SAM). The first is usually associated with patterned light beams, such as beams with Laguerre-Gaussian amplitude distribution \cite{Allen1992}, or displaced beams \cite{Aiello2009}. In contrast, SAM is associated with elliptical or circular polarization. \\
	In general, the SAM can be  quantified by the local spin density, which for light with frequency $\omega$ is defined as\cite{Bliokh2015a} 
	\begin{equation}
	\boldsymbol{S}=	\text{Im} [\epsilon_0 \boldsymbol{\mathcal{ E}}^* \times \boldsymbol{\mathcal{E}}+ \mu_0 \boldsymbol{\mathcal{H}}^* \times \boldsymbol{\mathcal{H}}]/ 4 \omega\;,	
	\label{eq:spindensity}		
	\end{equation}
	where $\epsilon_0$ ($ \mu_0$) is the vacuum permittivity (permeability) and $\boldsymbol{\mathcal{H}}$ is the complex vector amplitude of the magnetic field. In Eq.~(\ref{eq:spindensity}), we discriminate two contributions: The first part corresponds to the electric field induced spin density and the second to the magnetic field induced spin density. 
	For a circularly polarized plane wave propagating in $y$-direction, the electric as well as the magnetic fields are purely transverse, and their field vectors rotate in a plane perpendicular to $y$. Thus, the spin density points along the propagation direction and one speaks of a longitudinal SAM, i.e. $\boldsymbol{S_l}\!\equiv\!S_y\boldsymbol{e_y}\!\neq\! 0$ and $\boldsymbol{S_t}\!\equiv S_x\boldsymbol{e_x}+S_z\boldsymbol{e_z}\!\!=\! 0$ (see Fig.~\ref{fig:trans_vs_long_angular_momentum}a). From Eq.~(\ref{eq:spindensity}), we see that the transverse electric spin density always vanishes when we have only transverse field components, which is the case for a linearly polarized plane wave. However, for a wave that exhibits a longitudinal electric field component which has a phase difference to the transverse component, we obtain a non-vanishing transverse spin density vector, i.e. $|S_t|\!>\!0$ (see Fig.~\ref{fig:trans_vs_long_angular_momentum}b). Thus, the elliptical polarization which stems from the additional longitudinal electric field component gives rise to a transverse SAM \cite{Aiello2015,Banzer2013,Bliokh2015a}.\\
	In the framework of this chapter, we are mainly interested in coupling light to emitters that have strong electric dipole transitions. Thus, the properties of the electric field are of major concern to us and, in the following we limit our discussion to the electric part of the spin density. Nonetheless, the generalization to its magnetic counterpart is straightforward.

\subsection{Spin--momentum locking of light} 

The most important feature of transverse SAM is that it is inherently linked to the propagation direction of the light field. If we reverse the propagation direction, the sense of rotation of the field vector changes and the SAM flips its sign.
The inversion of propagation direction corresponds to changing the sign of the wave vector, i.e. $\boldsymbol{k}\to -\boldsymbol{k}$. From Eq.~(\ref{eqn2}), we see that this also changes the sign of the longitudinal field component and, thus, the rotation sense of the local elliptical polarization. For a wave propagating in positive $y$-direction ($\boldsymbol{k}\!>\!0$) with equal amplitudes for its longitudinal and transverse field components, we obtain perfectly circularly polarized light, $\boldsymbol{\mathcal{E}}_{\sigma^+}\!=\!|\boldsymbol{\mathcal{E}}|(\boldsymbol{e}_x+ i \boldsymbol{e}_y)/\sqrt{2}$ and thus $S_z>0$. In contrast, if we reverse the propagation direction ($\boldsymbol{k}\!<\!0$) the light has the orthogonal circular polarization, $\boldsymbol{\mathcal{E}}_{\sigma^-}\!=\!|\boldsymbol{\mathcal{E}}|(\boldsymbol{e}_x-i \boldsymbol{e}_y)/\sqrt{2}$ and the SAM points in the opposite direction, i.e. $S_z<0$. Thus, this effect is often referred to as spin-momentum locking (SML) of light. SML is a direct consequence of time reversal symmetry of Maxwell's equations. The time reversal operation, i.e. $t \mapsto -t$, is equivalent to inverting the wave vector, $\boldsymbol{k} \mapsto -\boldsymbol{k} $ and taking the complex conjugate of the electric field amplitude, $\boldsymbol{\mathcal{E}} \mapsto \boldsymbol{\mathcal{E}}^*$. Since $\boldsymbol{\mathcal{E}}_{\sigma^+}\!=\!(\boldsymbol{\mathcal{E}}_{\sigma^-} )^*$ this gives, for the case of transverse angular momentum, rise to the situation that the counter-propagating fields have orthogonal polarization.
\\

\subsection{Properties of WGMs}

In the previous section, we briefly discussed how transverse angular momentum and spin--momentum locking arises for light fields that strongly vary in transverse direction.
In the following, we want to see how these effects emerge in the case of WGM resonators. Such resonators are dielectric structures that provide radial confinement for light by means of total internal refection. In addition, a concave structure of the resonator prevents the light from escaping in axial direction. WGM resonators have been demonstrated in a large variety of geometries, including near cylinders\cite{Sumetsky2011}, prolate-shaped resonators\cite{Poellinger2009}, spheres\cite{Campillo:1991aa}, toroids\cite{Vahala:2003aa} and disks\cite{Lee2012a}. Guiding the light by total internal reflection gives rise to strong field gradients, in particular in the evanescent field, that can cause transverse SAM. In the following, we analyze this quantitatively by introducing an analytic model which allows us to calculate the electric field of WGMs.

\subsubsection{Analytic approximation in radial direction}

\begin{figure}
	\centering
		\includegraphics[width=0.8\columnwidth]{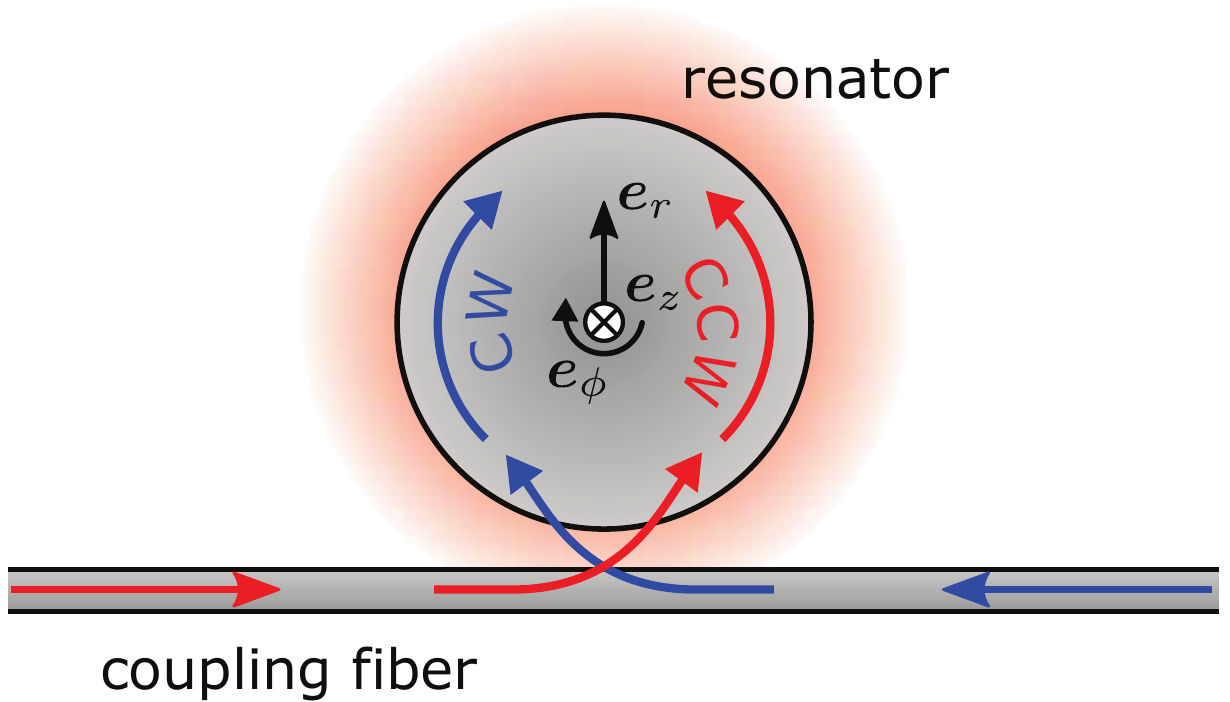}
	\caption{Schematic of a WGM resonator coupled to a waveguide. Light can either circulate in clockwise (CW) or counter clockwise (CCW) direction, in the $r$-$\phi$-plane, around the surface of the resonator. These modes can be excited by sending light through a coupling fiber in one or the other direction. }
	\label{fig:resonator_CW_CCW}
\end{figure}

In order to obtain an idea of the mode structure and the polarization properties of WGMs, we have to derive an accurate solution of the electromagnetic field for the resonator eigenmodes. Due to their spherical or cylindrical symmetry, there exists an exact analytic solution for spherical and nearly fully cylindrical resonators \cite{Oraevsky2002,Matsko:2006aa}. We will restrict our discussion to the latter case, since the result will be valid for the fundamental axial mode in the symmetry plane of any resonator geometry.\cite{Matsko:2006aa} To take advantage of the cylindrical symmetry, it is useful to change to cylindrical coordinates (radial $r$, azimuthal $\phi$, axial $z$), as shown in Fig.~\ref{fig:resonator_CW_CCW}. The electric field $\boldsymbol{\mathcal{E}}$ has to respect the Helmholtz equation
\begin{equation}
	 (\boldsymbol{\nabla}^2 + |\boldsymbol{k}|^2n^2(r)) \boldsymbol{\mathcal{E}}= 0 \, ,
\label{eq:PR_helmholtz}
\end{equation}
with the length of the light's wavevector $|\boldsymbol{k}|\!=\! 2\pi/\lambda \!=\! \omega/ c$, the vacuum wave length $\lambda$ and $n(r)$ the refractive index which is $n_0$ inside and, for simplicity, $n=1$  outside of the resonator. Equation~(\ref{eq:PR_helmholtz}) directly follows from Maxwell's equations in the absence of charges and currents. 
For a perfect cylinder, we can separate the propagation of the light into a propagation along the cylinder with wavevector $\boldsymbol{k}_z$ and around the cylinder's circumference with wavevector $\boldsymbol{k}_{\phi,r}$, where $\boldsymbol{k}=\boldsymbol{k}_{\phi,r}+\boldsymbol{k}_z$. This allows us to separate the Helmholtz equation using the Ansatz $\boldsymbol{\mathcal{E}}=\boldsymbol{\Phi}(r)\mathcal{Z}(z)\text{exp}(im\phi)$.
Here, we already inserted the solution of the azimuthal part which is defined by the azimuthal quantum number $m$. The radial differential equation now reads 
\begin{equation}
\partial^2_r \boldsymbol{\Phi}(r) +\frac{1}{r}\partial_r \boldsymbol{\Phi}(r)+\left(k^2_{\phi,r}-\frac{m^2}{r^2}\right)\boldsymbol{\Phi}(r)=0\;.
\end{equation}
The solutions are given by the Bessel function of the first kind $J_m$ and the Bessel function of the second kind $Y_m$.
Resonances of a cylindrical resonator with radius $R$ can then be found by solving the transcendental equation, that follows from Maxwell's electromagnetic boundary conditions\cite{Oraevsky2002}
\begin{equation}
P\frac{J'_m(k_{\phi,r}n_0R)}{J_m(k_{\phi,r}n_0R)}=\frac{Y'_m(k_{\phi,r}R)}{Y_m(k_{\phi,r}R)}\;,
\label{eq:transcendental}
\end{equation}
where the prime indicates a total derivative and $P=n_0$ or $1/n_0$ for the case when the electric field preferentially points parallel to the surface (TE) or when the electric field has a component perpendicular to the surface (TM), respectively.
An exact analytic solution of Eq.~(\ref{eq:transcendental}) is generally not possible, such that resonance frequencies must be determined either numerically or by means of analytical approximations.~\cite{Lam1992}
For a given radius $R$, the resulting component of the wave vector can be expressed as 
\begin{equation}
k_{\phi,r} =\frac{m\, f_p^m}{n_0\, R}\;,
\label{eq:kphi}
\end{equation}
where $f_p^m\sim1$ accounts for the geometric dispersion and $p$ is the radial quantum number that indicates the $p+1$ root of Eq.~(\ref{eq:transcendental}). 
The corresponding solutions of the radial wave equation inside and outside the resonator have the form
\begin{equation}
	\Phi(\phi,r)_l = \begin{cases}
		A_l\cdot J_m\left(n_0 rk_{\phi,r} \right), & \text{for  } r \le R \, , \\
		\\
		B_l\cdot Y_m\left(r k_{\phi,r}  \right), & \text{for  } r > R \, ,
	\end{cases}
\label{eq:PR_Bessel_solution}
\end{equation}
where the index $l$ represents the component of the electric field vector. To determine the relative magnitude of the constants $A_l$ and $B_l$, we again have to consider the boundary conditions for electromagnetic fields.
\label{sec:pol_bmr}

\subsubsection{SAM in WGMs}
\begin{figure*}
	\centerline{\includegraphics[width=7.5in]{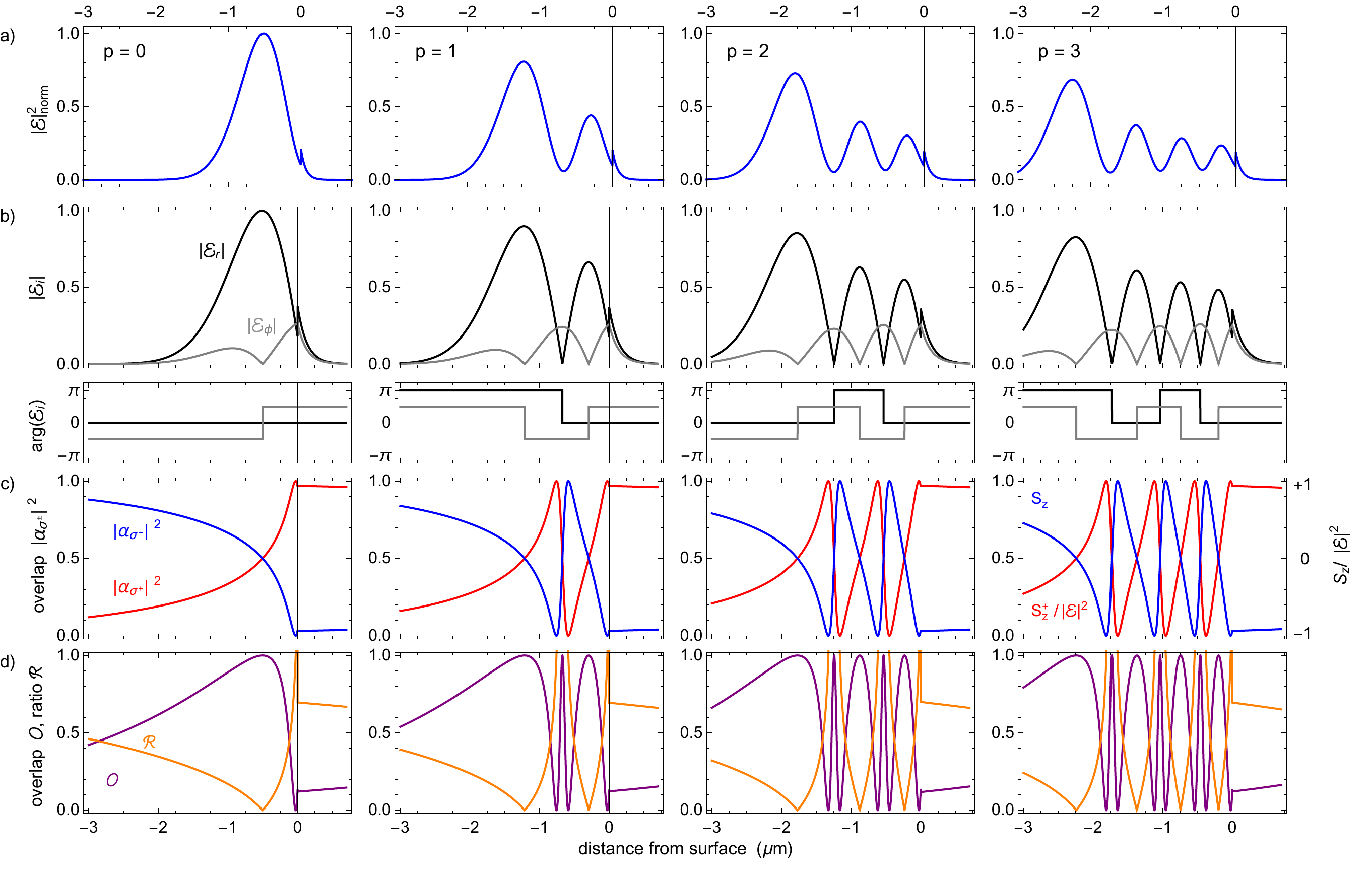}}
	\caption{Polarization properties of the electric field for the four lowest radial TM modes, $p=\{0,1,2,3\}$, calculated for $n_0=1.45$, $R=20~\mu$m and $m=206$. a) The intensity as a function of the distance from the resonator surface. b) The amplitude and phase of the transverse $\mathcal{E}_r$ (black) and longitudinal $\mathcal{E}_\phi$ (gray) electric field components. The electric fields are normalized to the maximum value of $|\mathcal{E}_\text{TM}|_{p=0}$. c) The polarization overlap of the local field of a counter-clockwise running wave mode with $\sigma^+$ (red) and $\sigma^-$ (blue) polarized light. Note that the normalized z-component of the spin density $S_z$ for the counter-clockwise running wave mode can be obtained via $S_z/|\mathcal{E}|^2=2|\alpha_{\sigma^+}|^2-1$. d) The polarization overlap $\mathcal{O}$ of two counter-propagating modes and the ratio between the longitudinal and the transverse field component $\mathcal{R}=|\mathcal{E}_\phi/\mathcal{E}_r|$. }
	\label{fig:radial_overla_spin_p0}
\end{figure*}

The solution given above accurately describes the electric field in the symmetry plane for any WGM resonator geometry. 
In Fig.~\ref{fig:radial_overla_spin_p0}a the normalized intensity distributions for the first four resonances $p\!=\!\{0,1,2,3\}$ are shown as a function of the distance from the surface in radial direction using the example of TM modes.
The radial quantum number $p$ gives the number of intensity minima in radial direction inside the dielectric $(r<0)$. Outside the dielectric $(r>0)$, the electric field decays exponentially, forming the evanescent field. Even though the intensity distribution inside the resonator changes drastically when varying $p$, the amplitude of the evanescent field is almost unaffected. A similar intensity distribution can also be obtained for TE modes.

We now take a closer look at the polarization properties of the modes.
For the case of TE-polarized modes we only get a single field component, that oscillates parallel to the resonator surface, i.e. transverse to the propagation direction along the circumference. Since along this direction no  field gradients occur, we expect no longitudinal field component and thus no transverse SAM. In contrast, for TM modes the field exhibits two polarization components, a transverse and a longitudinal one. 
Figure~\ref{fig:radial_overla_spin_p0}b shows the transverse, $\mathcal{E}_r$, and the longitudinal, $\mathcal{E}_\phi$, electric field component and their the respective phase as a function of the radial position for a TM polarized mode. While $\mathcal{E}_\phi$ is continuous at the surface, $\mathcal{E}_r$ is discontinuous. This gives rise to the step in the intensity distribution shown in Fig.~\ref{fig:radial_overla_spin_p0}a.

In order to describe the interaction between the resonator field and an emitter, such as a single atom, the local polarization overlap between the field and the eigenpolarization of the atom has to be determined.
Therefore, we introduce the complex valued field overlap between the polarization eigenstates $i \in \{\sigma_+,\pi,\sigma_-\}$ and the polarization of the resonator fields propagating in counter-clockwise (CCW) or clockwise (CW) direction, as
\begin{eqnarray}
	\alpha_{i} = \frac{\boldsymbol{\mathcal{E}}^+ \cdot \boldsymbol{e}^{\, *}_{i} }{|\boldsymbol{\mathcal{E}}^+|}  \hspace{0.5cm} \text{and}\hspace{0.5cm}
		\beta_{i} = \frac{\boldsymbol{\mathcal{E}}^- \cdot \boldsymbol{e}^{\, *}_{i} }{|\boldsymbol{\mathcal{E}}^-|} .
\label{eq:TN_alpha_beta}
\end{eqnarray}
Here, we use the electric field vector $\boldsymbol{\mathcal{E}}^\pm$ for waves that travel in $\pm \phi$-direction, i.e. CCW and CW. In addition, we defined the unit vectors $\boldsymbol{e}_{\sigma^+}\!=\!(\boldsymbol{e}_r + i \boldsymbol{e}_\phi ) / \sqrt{2}$, $\boldsymbol{e}_{\sigma^-}\!=\! (\boldsymbol{e}_r - i \boldsymbol{e}_\phi ) / \sqrt{2}$ and $\boldsymbol{e}_{\pi}\!=\!\boldsymbol{e}_z $, using the cylindrical unit vectors $\boldsymbol{e}_{r}$, $\boldsymbol{e}_{\phi}$and  $\boldsymbol{e}_{z}$. In Fig.~\ref{fig:radial_overla_spin_p0}c the polarization overlaps $|\alpha_{\sigma^+}|^2=|\beta_{\sigma^-}|^2$ and $|\alpha_{\sigma^-}|^2=|\beta_{\sigma^+}|^2$ are plotted as a function of the radial position. 
Inside the dielectric, the mode shows a variety of polarization states, ranging from perfectly circular ($|\alpha_{\sigma^+}|^2=1$ or $|\alpha_{\sigma^-}|^2=1$), whenever the longitudinal and transverse field components are of equal size, to linear local polarization ($|\alpha_{\sigma^\pm}|^2=0.5$), if one component is zero. In the evanescent field, the longitudinal and the transverse field component decay on nearly the same length scale, and their ratio is approximately constant. For a resonator made out of silica the ratio close to the surface is $|\mathcal{E}_\phi/\mathcal{E}_r|\approx 0.7$, largely independent of the specific mode (see Fig.~\ref{fig:radial_overla_spin_p0}d). This is expected for total internal reflection at gracing incidence. For this case, the polarization overlap of the CCW propagating mode with perfectly $\sigma^+$-polarized light is $|\alpha_{\sigma^+}|=0.97$ and with $\sigma^-$-polarized light is $|\alpha_{\sigma^-}|=0.03$. In contrast, for light in the CW mode the evanescent field has almost  perfect $\sigma^-$ polarization, i.e. $|\beta_{\sigma^-}|=0.97$ and $|\beta_{\sigma^+}|=0.03$.\\
Again, we can associate a SAM with this elliptical polarization. For TM modes, the only non-vanishing electric spin component is $S_z$, which is orthogonal to the propagation direction of the running wave modes, given by $\boldsymbol{e}_\phi$. The normalized spin density $S_z/|\mathcal{E}|^2$ is plotted in Fig.~\ref{fig:radial_overla_spin_p0}c as a function of the distance to the surface.   \\

In order to quantify the degree of SML in the resonator modes, we define the local polarization overlap of two counter-propagating modes using
\begin{equation}
\label{eq:overlap}
\mathcal{O}= \frac{|\boldsymbol{\mathcal{E}}^+ \cdot (\boldsymbol{\mathcal{E}}^-)^*|^2 }{|\boldsymbol{\mathcal{E}}^+|^2|\boldsymbol{\mathcal{E}}^-|^2} =|\Sigma_i \,\alpha_i^{\textcolor{white}{*}} \, \beta_i^*|^2\;.
\end{equation} 
If two counter-propagating waves have orthogonal local polarization, this value becomes zero, indicating maximum correlation between propagation direction and polarization. Thus, $\mathcal{O}<1$ is a clear indication for SML. However, if both waves have the same local polarization their overlap is 1, and they exhibit no SML. 
When superimposing two counter-propagating modes of equal amplitude, they will partially interfere and the contrast of the intensity modulation is given by $I_{\text{max}}-I_\text{min}=\sqrt{\mathcal{O}}$.
The overlap of two counter-propagating modes is shown in Fig.~\ref{fig:radial_overla_spin_p0}d. For the evanescent field close to the surface we obtain for a silica WGM resonator $\mathcal{O}=0.1$. 
Thus, two counter-propagating TM modes are almost orthogonally polarized. Even when we excite a pair of counter-propagating modes, they do not form an intensity modulated standing wave, but a polarization gradient field \cite{Dalibard1989,Lett1989} where the local polarization is always linear and the direction of polarization rotates a full turn every wavelength.\\
In the following, we will discuss in detail how we can incorporate the polarization properties into the description of light--matter interaction and how these affect the nature of this interaction.

\section{Coupling between a single atom and WGMs}
\label{sec_couplingWGM}
\begin{figure}
	\centering
		\includegraphics[width=\columnwidth]{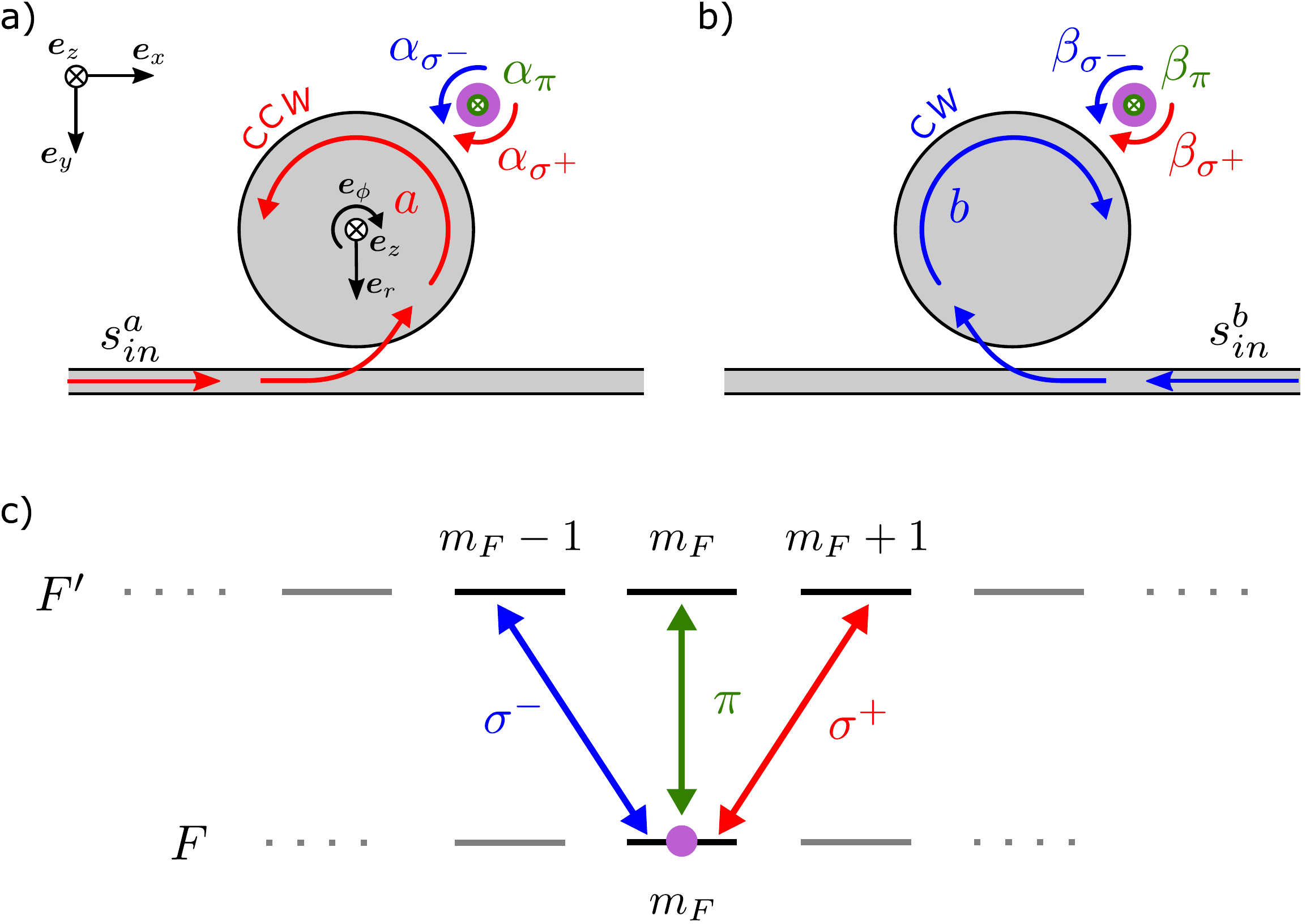}
	\caption{Schematic of light--atom coupling in WGM resonators. a) When launching light in the positive $x$-direction through a coupling waveguide, we excite the counter-clockwise (CCW) propagating running wave mode $a$. At the position of the atom, the polarization of the mode is described by the three field overlaps $\alpha_{i}=\boldsymbol{\mathcal{E}}^a \!\cdot \boldsymbol{e}_i^*/|\boldsymbol{\mathcal{E}}^a|$ between the local field vector $\boldsymbol{\mathcal{E}}^a $ and the atomic eigenpolarizations, $\boldsymbol{e}_i \in \{\boldsymbol{e}_{\sigma^-},\boldsymbol{e}_\pi,\boldsymbol{e}_{\sigma^+}\}$. b) When launching light in the negative $x$-direction through the coupling fiber we excite the clockwise (CW) propagating running wave mode $b$. At the position of the atom the polarization is described by the three field overlaps $\beta_{i}=\boldsymbol{\mathcal{E}}^b \!\cdot \boldsymbol{e}_i^*/|\boldsymbol{\mathcal{E}}^b|$ between the local field vector $\boldsymbol{\mathcal{E}}^b $ and the atomic eigenpolarizations. c) If the light in the resonator is resonant with the $F\!\rightarrow\!F'$ transition, the different polarization components $\{\sigma^-,\pi,\sigma^+\}$ drive to the excited state Zeeman levels $m_{F'}=m_F +\Delta_{m_F}$, with $\Delta_{m_F}=\{-1,0,+1\}$.}
	\label{fig:full_qm_model_sketch_level_scheme}
\end{figure}

The complex polarization structure of  WGMs and the resulting SML renders light-matter interaction in such resonators fundamentally different when compared to traditional Fabry-P\'{e}rot or ring-resonators. In previous works, these effect were not taken into account for the description of the interaction between light and matter and only recently it was shown that they can qualitatively change the nature of the coupling~\cite{Junge:2013bb} and give rise to chiral light--matter interaction.\\
In this section, we introduce a model based on the master equation approach that accurately describes the situation we encounter when coupling single atoms to WGMs. This model includes a vectorial treatment of the resonator fields, thus correctly describing SML in WGM resonators. Consequently, we also have to take into account the complex level structure of the coupled atoms or emitters. 
Based on this general model, we discuss a few exemplary cases for which we can derive analytic solutions and which can be applied to many realistic situations. This allows one to gain a deeper understanding of chiral light--matter interaction.\\

\subsection{Light--matter interaction in WGMs}
\label{sec:full_qm_theory}
The accurate description of the light--matter interaction in WGM resonators requires a quantum mechanical treatment, which takes into account the full vectorial nature of the resonator field. For the purpose of giving a general description, it is useful to decompose the polarization of the resonator modes into the eigenpolarization of the atom along its quantization axis. In the following, we consider two degenerate counter-propagating resonator modes, $a$ and $b$, which can, e.g., be pumped by sending light through the coupling waveguide from either direction (see Fig.~\ref{fig:full_qm_model_sketch_level_scheme}a-b).  Due to its internal structure, atoms are in general non-polarization maintaining scatterers, meaning that the atom--light coupling strength in general depends on the light's polarization and the internal atomic state. As a consequence, the two counter-propagating modes can couple to different transitions of the atom with different coupling strengths. In order to account for this, we have to consider the complete magnetic sub-structure of the atomic levels.\\
In the following, we consider the case where the resonator interacts with the $F \rightarrow F^\prime$ transition of a single atom, where $F$ and $F'$ are the total spin of the hyperfine ground and excited states, respectively. For sake of readability, we do not include the possibility that the atom has several ground or excited hyperfine states, although, our model can readily be extended to this case. The WGM with resonance frequency $\omega_R$ is probed using an external field with frequency $\omega_P$. The resonator-guided light can drive different atomic transitions, where $\pi$-polarized light will drive transitions with $\Delta m_F = 0$ and $\sigma^\pm$-polarized light drives transitions with $\Delta m_F = \pm 1$ (see Fig.~\ref{fig:full_qm_model_sketch_level_scheme}c). Here, $m_F$ and $m_{F^\prime}$ are the Zeeman sublevels of the ground and excited state, respectively. 
In a reference frame rotating at the probe light frequency $\omega_P$, the uncoupled parts of the Hamiltonian are given by
\begin{equation}
\begin{aligned}
\hat{H}_\text{A}/\hbar  &=  \sum_{m_F=-F}^{F} \mu_B g_F m_F B  \ket{F,m_F} \bra{F,m_F}  \\
+&\sum_{m_{F'}=-F'}^{F'} (\mu_B g_{F'} m_{F'} B+\Delta_{AP})  \ket{F',m_{F'}} \bra{F',m_{F'}},
\label{eq:TN_H0_multi_level1}
\end{aligned}
\end{equation}
and
\begin{equation}
\begin{aligned}
\hat{H}_\text{R}/\hbar \, &=  \Delta_{RP}( \hat a^\dagger \hat a +  \hat b^\dagger b ) \, .
\label{eq:TN_H0_multi_level2}
\end{aligned}
\end{equation}
Here, we introduced $|F,m_F\rangle$ and $|F',m_{F'}\rangle$ as the atomic ground and excited states, respectively, and $\Delta_{AP}=\omega_A-\omega_P$ ($\Delta_{RP}=\omega_R-\omega_P$) is the detuning between the atomic transition (resonator) and the pump field. We also assume the presence of a magnetic field $B$ that defines the quantization axis and results in an energy shift for each Zeeman level of $\Delta E_Z / \hbar\!=\! \mu_B g_F m_F B$, where $\mu_B$ is the Bohr magneton and $g_F$ is the Land\'e $g$-factor of the corresponding energy level. $\hat{a}$ and $\hat{b}$ ($\hat{a}^\dagger$ and $\hat{b}^\dagger$) are the photon annihilation (creation) operators of the corresponding resonator mode.
The interaction part of the Hamiltonian is given by
\begin{equation}
\hat{H}_\text{I}/\hbar = g_a \hat a \hat d_a^\dagger + g_a^* \hat a^\dagger \hat d_a  + g_b \hat b \hat d_b^\dagger + g_b^* \hat b^\dagger \hat d_b  \, .
\label{eq:TN_Hint_multi_level}
\end{equation}
Here, $g_a$ and $g_b$ describe the interaction strengths of the atom with the two resonator modes. More specifically, $g_i= \psi_i(\boldsymbol{r}) \sqrt{\omega_R / 2\hbar\epsilon_0 V}\,\tilde \mu$,
where $\psi_i(\boldsymbol{r)}$ is the normalized mode function of the respective resonator field at the position of the atom $\boldsymbol{r}$ with the mode volume $V=\int{|\psi_i(\boldsymbol{r})|^2d^3r}$ of the resonator mode,
and $\tilde \mu=\sqrt{(2J+1)/(2J'+1)} \cdot \langle J||e\hat{\boldsymbol{r}}||J'\rangle$ is the reduced atomic transition dipole matrix element~\cite{Steck2001} with the dipole operator $e\hat{\boldsymbol{r}}$ and the total angular momenta $J$ and $J'$ of the ground and excited states, respectively. The total atomic lowering operators for the interaction with the two resonator modes can be separated into the atomic eigenpolarizations and are then given by
\begin{equation}
\begin{aligned}
\hat d_a &=& \alpha_{\sigma^+} \hat d_{+1} &+\alpha_{\pi} \hat d_{0}+\alpha_{\sigma^-} \hat d_{-1} \, ,\\
\hat d_b &=& \beta_{\sigma^+} \hat d_{+1} &+\beta_{\pi} \hat d_{0}+\beta_{\sigma^-} \hat d_{-1} \, ,
\label{eq:TN_lowering_b}
\end{aligned}
\end{equation}
respectively. In this expression, the complex coefficients $\alpha_i$ and $\beta_i$ reflect the polarization overlap between the atomic eigenpolarizations and the field of the resonator modes $a$ and $b$, which are defined in Eq.~(\ref{eq:TN_alpha_beta}). The operators $\hat d_{\Delta m_F}$ are the atomic lowering operators that include all possible transitions with $\Delta m_F \in \{-1,0,+1\}$ and are given by\footnote{Note that we added the term $\sqrt{(2J'+1)/(2J+1)}$ in accordance with Ref. \cite{SteckQAO}. Consequently, $g_i$ corresponds to the coupling strength for driving the cycling transition. }
\begin{equation}
\begin{aligned}
&\hat d_{\Delta m_F} =\\
& \sqrt{\frac{2J'+1}{2J+1}}\sum_{m_F=-F}^{F}  \mu_{m_F}^{m_F+\Delta m_F} \ket{F,m_F}  \bra{F',m_F+\Delta m_F} .
\end{aligned}
\end{equation}
The relative strengths of the respective transitions can be expressed as~\cite{Steck2001}
\begin{eqnarray}
\mu_{m_F}^{m_{F'}}&=&\sqrt{(2F'+1)(2F+1)(2J+1)}\nonumber\\
&&\times \begin{pmatrix}
F' & 1 & F\\
m_{F'} & q & -m_F
\end{pmatrix}
\left\lbrace \begin{matrix}
J & J' & 1\\
F' & F & I
\end{matrix}\right\rbrace.
\label{eq:dipolmatrixelement}
\end{eqnarray}
Here, $q=(0,\pm1)$ for transition involving $(\pi,\sigma^\pm)$ polarized light and the last two terms are the Wigner 3-j and 6-j symbols, respectively.\\
For completeness, we also introduce a coupling rate $h$ between the resonator modes, due to scattering by e.g. surface roughnesses or impurities. The full Hamiltonian of the coupled atom--resonator system finally reads
\begin{eqnarray}
\hat{H} / \hbar &=& (\hat{H}_\text{A}+\hat{H}_\text{R}+ \hat{H}_\text{I})/ \hbar +  h (\hat a^\dagger +\hat b^\dagger)( \hat a+\hat b) \nonumber\\
&&+ i \epsilon_a (\hat{a}-\hat{a}^\dagger)+ i \epsilon_b (\hat{b}-\hat{b}^\dagger) .
\label{eq:TN_H_wgm}
\end{eqnarray}
Here, the last two terms describe the pumping of the resonator modes $a$ and $b$ by external light fields via the coupling fiber. For the case where the incident wave can be treated as a classical coherent field that is coupled into the resonator via the waveguide, $\epsilon_a\approx\sqrt{2\kappa_{ext}}\langle s^{a}_{in}\rangle$ and $\epsilon_b\approx\sqrt{2\kappa_{ext}}\langle s^{b}_\text{in}\rangle $, where $\langle s^{a}_\text{in}\rangle$ and $\langle s^{b}_\text{in}\rangle$ is the mean amplitude of the field propagating in positive or negative $x$-direction through the waveguide \cite{Peano2010} and $\kappa_{ext}$ is the waveguide-resonator coupling rate. The full time evolution, including decoherence, can be calculated using the master equation approach, where we have to take into account all the decay channels including the atomic excited state, as well as, losses from and both counter-propagating modes
\begin{eqnarray}
\frac{d \hat{\rho}}{dt} &=& -\frac{i}{\hbar} [\hat{H},\hat{\rho}] + (\kappa_{ext}+\kappa_0) (\mathcal{D}[\hat a] + \mathcal{D}[\hat b])\nonumber\\
&&  +  2\gamma \sum_{m_F,m_{F'} } \mathcal{D}[\hat d^{m_{F'}}_{ m_F}]  \, ,
\label{eq:TN_master_equation_wgm}
\end{eqnarray}
where we have introduced
\begin{equation}
\mathcal{D}[\hat c] = 2\hat c\rho \hat c^\dagger-\hat c^\dagger  \hat c \rho-\rho  \hat c^\dagger \hat c \, 
\label{eq:TN_Superoperator}
\end{equation}
and $\gamma$ ($\kappa_0$) is the atomic (intrinsic resonator) field decay rate. If several ground or excited hyperfine states have to be considered, the sums in Eq.~(\ref{eq:TN_H0_multi_level1}) and Eq.~(\ref{eq:TN_master_equation_wgm}) have to be extended to all possible states.\\

In order to derive the steady state response of the system, we can solve the full master equation for the case $\dot{\rho}=0$ and extract the expectation values of the resonator field amplitudes $\braket{\hat{a}}$ and $\braket{\hat{b}}$.  When we probe the system only from one side such that we drive mode $a$ we can then calculate the fiber transmission amplitude from the input--output relation 
\begin{equation}
\hat{s}^a_\text{out}= \hat{s}^a_\text{in}-i \sqrt{2\kappa_\text{ext}}\,\hat{a}\;,
\label{eq:out_transmission}
\end{equation}
via
\begin{equation}
\label{eq:tran_equ}
t = 1-i \sqrt{2\kappa_\text{ext}}\frac{\braket{\hat{a}}}{\braket{s^a_\text{in}}} \;.
\end{equation}
Due to direct mode--mode coupling and scattering of light between the two counter-propagating modes via the atom, mode $b$ can also become populated. The light in mode $b$ can couple back into the coupling fiber, leading to a finite refection signal, that can be calculated using
\begin{equation}
\hat{s}^b_\text{out}= -i \sqrt{2\kappa_\text{ext}}\,\hat{b}\;,
\label{eq:out_reflection}
\end{equation}
 via
\begin{equation}
\label{eq:ref_equ}
r = -i \sqrt{2\kappa_\text{ext}}\frac{\braket{\hat{b}}}{\braket{s^a_\text{in}}}\;.
\end{equation}
When probing from the opposite direction, i.e. pumping mode $b$, the mode operators $\hat{a}$ and $\hat{b}$ have to be interchanged.\\

\subsection{Simple analytic solutions}

In order to develop a clear understanding about the consequences of SML of the resonator modes in conjunction with the multi-level structure of the atom, we will now discuss three simple but instructive examples.
We assume the atom to be located close to the symmetry plane ($z=0$) and align its quantization axis with the symmetry axis of the resonator ($z$-axis). Consequently,  the polarizations of TM and TE modes coincide with the atomic eigenpolarizations $\boldsymbol{e}_{\sigma^\pm}$ and $\boldsymbol{e}_{\pi}$. Moreover, time reversal symmetry of Maxwell's equations implies that the field overlaps fulfill $|\alpha_{\sigma^+}|\!=\!|\beta_{\sigma^-}|$,  $|\alpha_{\sigma^-} |\!=\!|\beta_{\sigma^+}|$ and $|\alpha_{\sigma^-}|^2\!=\!1\!-\!|\alpha_{\sigma^+}|^2$. In this basis, the atomic operators simplify to $\hat d_a =\hat d_b= \hat d_{0}$ for TE modes and to $\hat d_a = \alpha_{\sigma^+} \hat d_{+1}+\alpha_{\sigma^-} \hat d_{-1}$ and $\hat d_b = \beta_{\sigma^+} \hat d_{+1}+ \beta_{\sigma^-} \hat d_{-1}$ for TM modes. Furthermore, both running wave modes $a$ and $b$ have the same field amplitude at the position of the atom, and thus we get $g_a=g_b=g$.\\
Until now, we made no approximation and this model provides the exact description of the interaction of a single scatterer with WGM resonator modes. Let us now consider the case where the atom is prepared in a well defined ground state $\ket{F,m_\text{F}}$. For TM-modes, the light in the resonator modes can now only drive the $\Delta m_\text{F}=\pm 1$ transitions, as depicted in Fig.~\ref{fig:full_qm_model_sketch_level_scheme}c. The respective interaction strength of the two competing transitions  depends on two parameters. The first is the polarization overlap of the local field with $\sigma^+$or $\sigma^-$. The second are the dipole matrix elements for the respective transitions. \\
In the following, we will discuss three examples, where we  limit the number of atomic levels involved in the interaction. First, we consider the symmetric case, where the scatterer equally couples to $\sigma^+$ and $\sigma^-$-polarized light. Secondly, we study the case where the scatterer exclusively couples to $\sigma^+$-polarized light. For completeness, we will then consider the case where the scatterer exclusively couples to linearly polarized light. 
For all three cases we will be able to derive an analytic solution for the steady state of the cavity modes $\braket{a}$ and~$\braket{b}$, that will be presented in the following.

\subsubsection{Coupling between polarization-independent scatterer and WGMs}
\begin{figure*}
	\centering
		\includegraphics[width=0.9\textwidth]{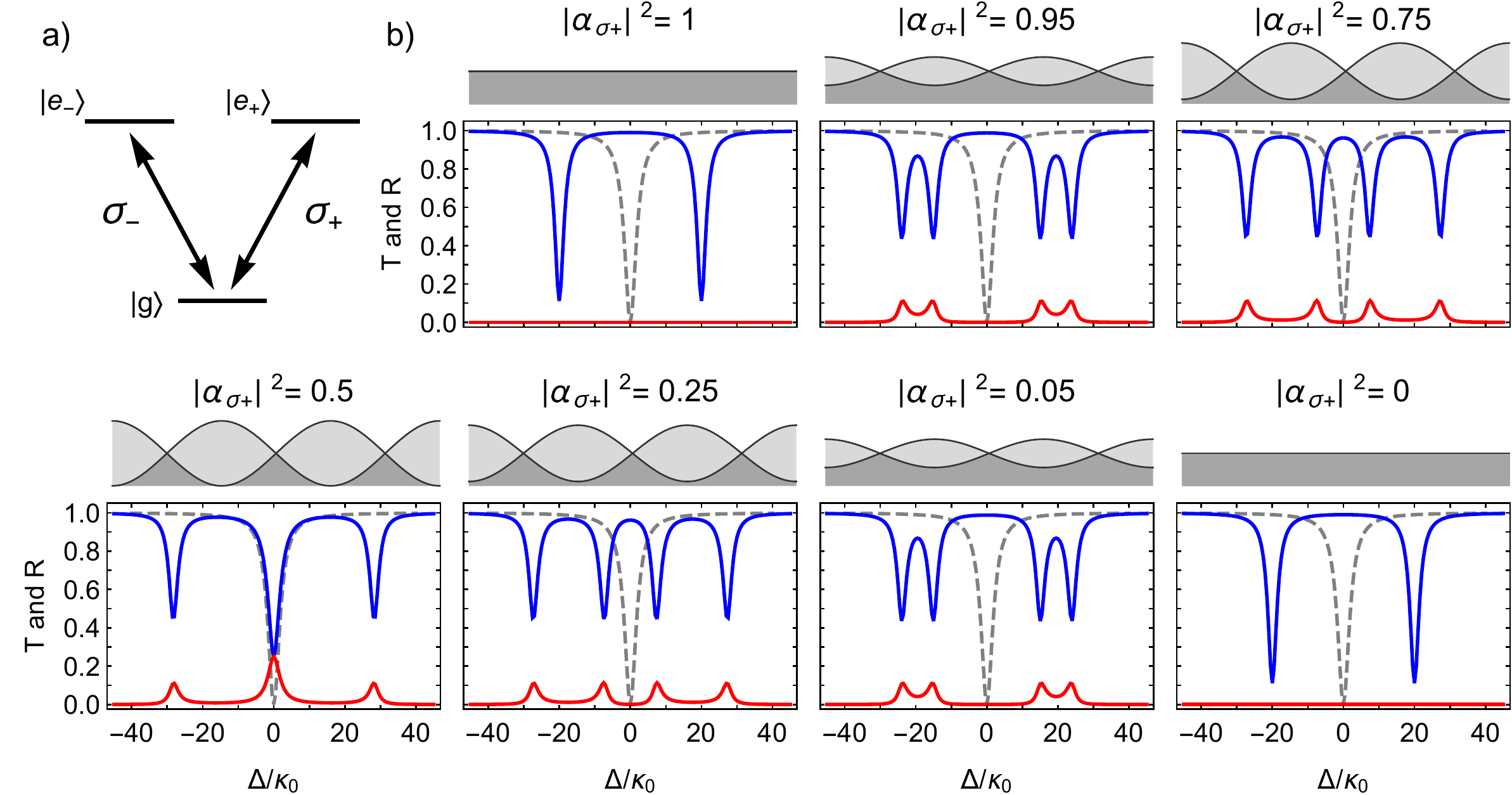}
	\caption{a) V-like level structure, realized e.g. in a $\ket{F\!=\!0 }\rightarrow \ket{F'\!=\!1} $ transition, with a quantization axis that is aligned with the resonator. The atom is coupled to TM polarized light, which can drive $\sigma^+$ and $\sigma^-$ transitions. For this level structure, the coupling between the atom and $\sigma^+$- or $\sigma^-$-polarized light is the same. b) Transmission (blue) and reflection (red) through the waveguide coupled to the atom--resonator system, for different overlaps $|\alpha_{\sigma^+}|^2$. Due to SML of the resonator field, the change from $|\alpha_{\sigma^+}|^2>1/2$ to $|\alpha_{\sigma^+}|^2<1/2$ can be interpreted as a change of the probing direction. For comparison, the empty resonator spectrum is plotted as a dashed line. The small inset on top of each spectrum represents the intensity modulation of two standing wave modes for the respective setting. All shown spectra were calculated using Eq.~(\ref{eq:modes_ab_3la}) and the following parameters: $(g,\gamma,\kappa_\text{ext},h)=(20,1,1,0)\cdot\kappa_0$. For these parameters the system is in the strong coupling regime of CQED, i.e. $g\gg (\kappa_0,\kappa_\text{ext},\gamma)$ }
	\label{fig:Spectra_3la}
\end{figure*}
Symmetric coupling can be, e.g., realized when the atom is prepared in the $\ket{F\!=\!0,m_\text{F}\!=\!0}$ ground state. The TM-polarized light in the resonator modes now drives $\Delta m_\text{F}=\pm1$ transitions between the ground state and the $\ket{F'\!=\!1,m_\text{F'}\!=\!\pm1}$ excited states, forming a V-like level structure, as depicted in Fig.~\ref{fig:Spectra_3la}a. For these two transitions, the dipole matrix elements are of equal size, $\mu_{0}^{+1}= \mu_{0}^{-1}$. Therefore, the total interaction strength between the light field and the atom does not depend on the light's polarization. Thus, the atom couples to TM modes independent of the light's propagation direction. However, depending on its polarization, the light in the resonator couples to one or both atomic  transitions. This already enables us to derive an analytic steady state solution for the expectation value of the two cavity modes. In the absence of mode--mode coupling and if only mode $a$ is pumped we obtain
\begin{align}
\label{eq:modes_ab_3la}
\braket{\hat{a}}&=
\frac{-i \sqrt{2 \kappa _{\text{ext}}}\tilde{\gamma}  \, \braket{s^a_\text{in}}}{\tilde{\gamma} \tilde{\kappa} (2g^2+\tilde{\gamma} \tilde{\kappa})+g^4 \left(2|\alpha_{\sigma^+}|^2-1\right)^2}\left(g^2  + \tilde{\gamma} \tilde{\kappa} \right)\;,\\	
	\braket{\hat{b}}&=
	\frac{i \sqrt{2 \kappa _{\text{ext}}}  \tilde{\gamma}\, \braket{ s^a_\text{in}}}{\tilde{\gamma} \tilde{\kappa} (2g^2+\tilde{\gamma} \tilde{\kappa})+g^4 \left(2|\alpha_{\sigma^+}|^2-1\right)^2}\;2\;g^2\alpha_{\sigma^+}^*\beta_{\sigma^+}\,\;.
	\label{eq:modes_ab_3lb}
  \end{align} 
Here, the atom and resonator detunings have been absorbed into the new variables, $\tilde{\gamma}=\gamma +i \Delta_{AP} $ and $\tilde{\kappa} =\kappa _{\text{ext}}+\kappa _0+i \Delta_{RP}$. Equations~(\ref{eq:modes_ab_3la}) and (\ref{eq:modes_ab_3lb}) not only contain the usual system parameters, $g, \tilde \kappa$ and $\tilde{\gamma}$, but also depend on the polarization overlap of the atomic dipole with the local fields via $\alpha_{\sigma^+}$ and $\beta_{\sigma^+}$.\\
\begin{figure*}
	\centering
	\includegraphics[width=0.9\textwidth]{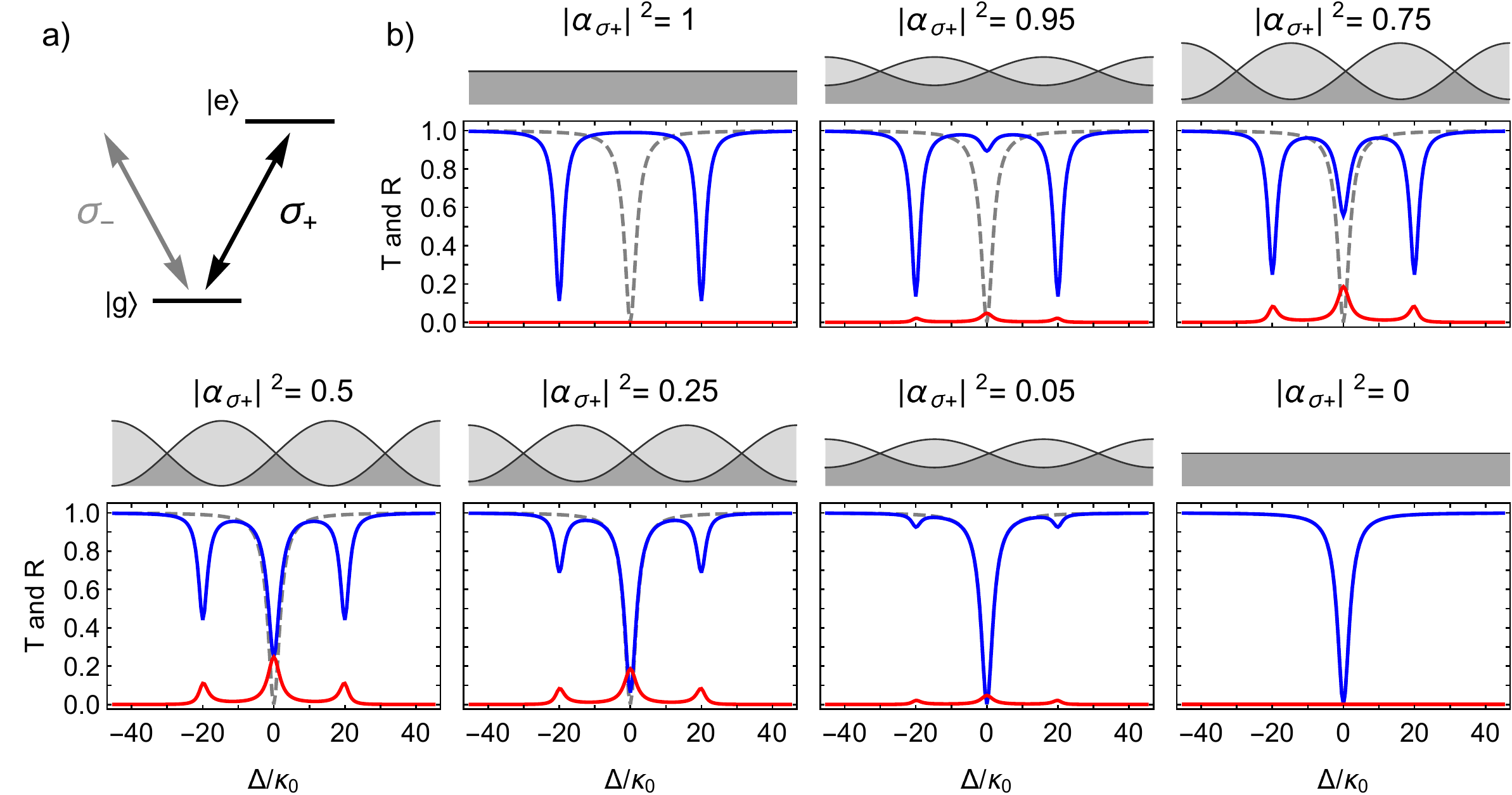}
	\caption{a) Same as Fig.~\ref{fig:Spectra_3la} but for an effective two level atom, where only $\sigma^+$-polarized light couples to the atom. b) Transmission (blue) and reflection (red) through the waveguide coupled to the atom--resonator system, for different overlaps $|\alpha_{\sigma^+}|^2$. For comparison, the empty resonator spectrum is plotted as a dashed line. The small inset on top of each spectrum represents the intensity modulation of two standing wave modes for the respective settings. All shown spectra were calculated using Eq.~(\ref{eq:modes_ab_2la}) and the same parameters as for Fig.~\ref{fig:Spectra_3la}.}
	\label{fig:Spectra_2la}
\end{figure*}
Figure~\ref{fig:Spectra_3la}b, shows the transmission and reflection spectra measured through the coupling waveguide, calculated using Eqs.~(\ref{eq:tran_equ}-\ref{eq:ref_equ}) together with Eqs.~(\ref{eq:modes_ab_3la}-\ref{eq:modes_ab_3lb}), for different polarization overlap. 
If the resonator field is perfectly circularly polarized, i.e $|\alpha_{\sigma^+}|^2\!=\!1$ or $|\alpha_{\sigma^+}|^2\!=\!0$, it drives a closed cycling transition between the ground state and one of the two excited states. Since there is no polarization overlap between the two counter-propagating modes, no light will be scattered from mode $a$ into mode $b$ via the atom. This realizes the ideal situation where each resonator mode is independently coupled to an effective two-level atom. The transmission spectra shows two resonances separated by $2g$, and the reflection is zero for all detunings. If the polarization deviates from perfect circular polarization, i.e. $|\alpha_{\sigma^+}|^2\!\neq\! (0,1)$ the situation drastically changes. Instead of a two-resonance spectrum, we observe two additional resonances emerging. In order to understand the physical origin of this behavior, it is instructive to change basis of the resonator modes and consider the standing wave modes defined by the operators  $\hat A\!=\!(\hat a+\hat b)/\sqrt{2}$ and $\hat B\!=\!(\hat a-\hat b)/\sqrt{2}$. When there is a finite polarization overlap between the two running wave modes, the standing wave modes $\hat A $ and $\hat B$ will show intensity modulations along the circumference. This modulation is schematically shown as inset above each spectrum in Fig.~\ref{fig:Spectra_3la}b. Without loss of generality, we assume the atom to be at a maximum of one standing wave mode, and thus at a minimum of the other mode. Since the coupling strength of the atom to the resonator depends on the local field strength, it differs for the two standing wave modes and the spectrum shows two pairs of resonances. If the polarization overlap between the two running wave modes is increased, the intensity modulation becomes further pronounced and the two central resonances will move closer together while the two outer resonance slightly move apart. For $|\alpha_{\sigma^+}|^2\!=\!1/2$, both modes, $a$ and $b$, have the same linear polarization, i.e. are in an equal superposition of $\sigma^+$- and $\sigma^-$-polarized light. As a consequence, we obtain two fully modulated standing waves, of which only one couples to the atom. In this case, the two central resonances will merge at zero detuning, while the outer dips show an enlarged splitting of $2\sqrt{2}g$. This is caused by the intensity modulation which reduces the effective mode volume of the standing wave modes. 
However, for this setting, only half of the light that couples from the fiber to the resonator actually interacts with the atom. As direct consequence, the on-resonance transmission is bound to 0.25, for the particularly interesting case of strong coupling and impedance matched waveguide, i.e. $\kappa_\text{ext}=\kappa_0$. In addition, the atom will scatter light between the modes, which populates the un-pumped mode $b$ and results in a finite reflection. Note that this is typically the situation encountered in macroscopic ring resonators. \\
When we change the sign of the spin of the local polarization, the system responds symmetrically, meaning that we obtain the same spectra (cf. the first and the last panel in Fig.~\ref{fig:Spectra_3la}b). Due to SML, this situation corresponds to sending light from the opposite direction through the waveguide and thus directly pumping the mode $b$. \\

\subsubsection{Coupling between polarization-dependent scatterer and WGMs}
\label{sec:tla}
\begin{figure*}
	\centering
	\includegraphics[width=0.9\textwidth]{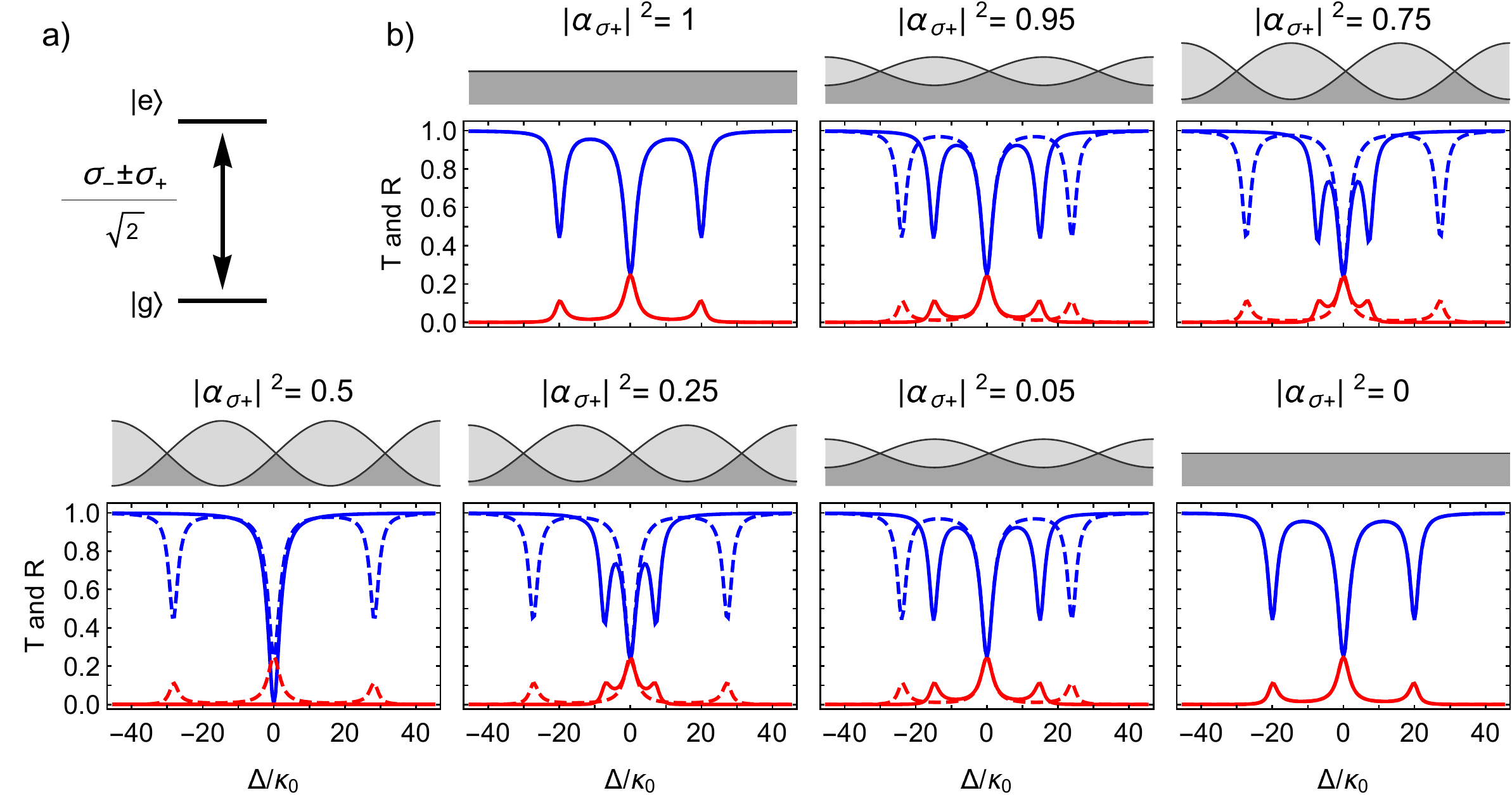}
	\caption{a) Same as Fig.~\ref{fig:Spectra_3la} but for a linearly polarized two level system. b) Transmission (blue) and reflection (red) through the waveguide coupled to the atom--resonator system, for different overlaps $|\alpha_{\sigma^+}|^2$. The dipole of the scatterer is aligned along $\boldsymbol {e}_r=(\boldsymbol{e}_{\sigma^+}+\boldsymbol{e}_{\sigma^-})/\sqrt{2}$ (dashed line) or $\boldsymbol {e}_\phi=(\boldsymbol{e}_{\sigma^+}-\boldsymbol{e}_{\sigma^-})/\sqrt{2}$ (solid line). For $|\alpha_{\sigma^+}|^2=1/2$, this yields maximal or zero coupling, respectively. The small inset on top of each spectra represents the intensity modulation of the two standing wave modes for the respective settings. All shown spectra were calculated using Eq.~(\ref{eq:modes_ab_2lb_lin}) and the same parameters as for Fig.~\ref{fig:Spectra_3la}.}
	\label{fig:Spectra_lintla}
\end{figure*}
Let us now consider the case where the scatterer exclusively couples to $\sigma^+$-polarized light. Such a situation can, e.g., be obtained by preparing an alkali atom in its outermost $m_\text{F}$-state, such that $\sigma^+$-polarized light drives a closed cycling transiting between $\ket{F,m_\text{F} }$ and $ \ket{F',m_\text{F}+1}$. In this case, the transition strength of the competing transition, $\ket{F,m_\text{F}} \rightarrow \ket{F',m_\text{F}-1}$, which is driven by $\sigma^-$-polarized light, is significantly smaller and can typically be neglected.
Thus, the level scheme reduces to an effective two-level system, which exclusively couples to $\sigma^+$-polarized light, as shown in Fig.~\ref{fig:Spectra_2la}a. For this case, the analytic solutions for the steady state of the resonator modes are given by
\begin{align}
\label{eq:modes_ab_2la}
\braket{\hat{a}}&=\frac{-i\sqrt{2\kappa _{\text{ext}}}  \, \braket{s_\text{in}^a}}{\tilde{\kappa} \left(g^2 +\tilde{\gamma} \tilde{\kappa}\right)}\,\left((1-|\alpha_{\sigma^+}|^2) g^2 + \tilde{\gamma} \tilde{\kappa}\right) \;,\\	
\label{eq:modes_ab_2lb}
	\braket{\hat{b}}&=\frac{i\,\sqrt{2\kappa _{\text{ext}}}  \, \braket{s_\text{in}^a}}{\tilde{\kappa}
   \left(g^2+\tilde{\gamma} \tilde{\kappa}\right)}\, g^2  \,\alpha_{\sigma^+}^*\beta_{\sigma^+}\;.
  \end{align} 
By inserting these formulas into Eqs.~(\ref{eq:tran_equ})-(\ref{eq:ref_equ}), we can again calculate the transmission and reflection through the waveguide, which are plotted in Fig.~\ref{fig:Spectra_2la}b.\\
If the local polarization of the pumped mode perfectly coincides with $\sigma^+$, i.e. $|\alpha_{\sigma^+}|^2=1$, the light in mode $a$ drives the atomic transition, while light in mode $b$ does not interact with the atom. The spectral response is the same as that which we obtained for the polarization independent scatterer and shows a two-resonance spectrum, where the resonances are split by $2g$. When the overlap of the mode with $\sigma^-$-polarized light is small but finite, i.e. $|\alpha_{\sigma^+}|^2\approx1$, a part of the light does not interact with the atom, causing an additional small transmission dip emerging on resonance. At the same time, the depth of the resonances at $\pm g$ decreases.  For $|\alpha_{\sigma^+}|^2=1/2$ and critical coupling to the empty resonator ($\kappa_\text{ext}=\kappa_0$), the on-resonant transmission again reaches 0.25. For $|\alpha_{\sigma^+}|^2=0$, the light does not couple to the atom any more and the spectrum resembles that of an empty resonator. In contrast to the previous model, the position of the resonance dips at $\Delta=0,\pm g$ is independent of $|\alpha_{\sigma^+}|^2$ and only their depths change. \\
When we change the probing direction through the coupling fiber, we drive the other of the two counter-propagating WGMs. 
Due to SML in TM modes, the two counter-propagating resonator modes have different elliptical polarization. In combination with a scatterer that only couples to $\sigma^+$-polarized light, this realizes a highly asymmetric transmission through the waveguide (cf. the first and the last panel in Fig.~\ref{fig:Spectra_2la}b). This effect can be employed to create nonreciprocal devices, such as optical diodes \cite{Sayrin2015bb} or optical circulators \cite{Scheucher2016} as discussed in section \ref{chap:diode}.

\subsubsection{Coupling between linearly polarized scatterer and WGMs}
\label{sec:lintla}

To complete the discussion of possible scatterers, we now discuss a linearly polarized scatterer that is aligned such that it couples to an equal superposition of $\sigma^+$ and $\sigma^-$-polarized light.
For this case, the two counter-propagating modes couple with equal strength to the scatterer, only the absolute coupling strength changes with varying polarization of the modes. The analytic solutions for the steady state of the resonator modes are

\begin{align}
\label{eq:modes_ab_2la_lin}
\braket{\hat{a}}&=\frac{-i\sqrt{2\kappa _{\text{ext}}}\braket{s_\text{in}^a}}{\tilde{\kappa} \left(2g^2|\alpha_{\text{lin}}|^2 +\tilde{\gamma} \tilde{\kappa}\right)}\, \left(|\alpha_{\text{lin}}|^2 g^2 + \tilde{\gamma} \tilde{\kappa}\right) \;,\\	
\label{eq:modes_ab_2lb_lin}
	\braket{\hat{b}}&=\frac{i\,\sqrt{2\kappa _{\text{ext}}} \,\braket{s_\text{in}^a}}{\tilde{\kappa}
   \left(2g^2|\alpha_{\text{lin}}|^2+\tilde{\gamma} \tilde{\kappa}\right)}\, g^2  \,\alpha_{\text{lin}}^*\beta_{\text{lin}}\;.
  \end{align} 
Here, we use the field overlap $\alpha_\text{lin}$ and $\beta_\text{lin}$ of the counter-clockwise and clockwise propagating mode with the linear eigenpolarization of the atomic transition. Note that $|\alpha_\text{lin}|=|\beta_\text{lin}|$. 
Figure~\ref{fig:Spectra_lintla} shows the calculated transmission and reflection spectra for an emitter aligned along $\boldsymbol {e}_r=(\boldsymbol{e}_{\sigma^+}+\boldsymbol{e}_{\sigma^-})/\sqrt{2}$ or along the azimuthal direction $\boldsymbol{e}_\phi=(\boldsymbol{e}_{\sigma^+}-\boldsymbol{e}_{\sigma^-})/\sqrt{2}$. In contrast to the previous systems, the emitter always couples equally to both counter-propagating modes. As a consequence, only half of the light interacts with the scatterer, yielding three resonances in the spectra. Note that, when the polarization of the WGMs or the orientation of the scatterer are changed Eq.~(\ref{eq:modes_ab_2la_lin}) and (\ref{eq:modes_ab_2lb_lin}) stay valid, only the effective coupling strength between the modes and the scatterer and therefore the resonance splitting changes ( see Fig.~\ref{fig:Spectra_lintla}). For $|\alpha_{\sigma^+}|^2=1$ or 0 both orientations show the same spectra: We observe resonance dips at $\Delta=0,\pm g$. When the overlap $|\alpha_{\sigma^+}|^2=1/2$, the longitudinal field component becomes zero. Thus, for a scatterer aligned along $\boldsymbol {e}_r$ we obtain maximal coupling and the resonance dips are at $\Delta=0,\pm \sqrt{2}g$. In contrast, for a fully misaligned scatterer, i.e. aligned along $\boldsymbol {e}_\phi$, the coupling becomes zero and only the empty resonator dip remains in the spectrum.
Even tough we considered resonator modes exhibiting SML, this system has fully symmetric transmission properties and does not show any chiral effects.

\section{Chiral waveguides}
\label{sec:chiral_wg}

In the last section, we saw that strongly coupling a single atom to a WGM can give rise to a direction-dependent, i.e. chiral, interaction between light and matter. For most applications it is desirable to deterministically couple photons to matter, and then extract the photon again with high probability. This requires one to operate the experiments in the so-called \textit{fast cavity} or \textit{Purcell} regime \cite{Reiserer2015}. In this regime, the coupling between the waveguide and the resonator is the dominant rate, while all actual losses are ideally smaller than the coherent emitter-light interaction $g$, i.e.  
\begin{equation}
\kappa_\text{ext}>g > \kappa_0,\gamma  \;.
\end{equation}
In many experimental settings, the coupled atom--resonator--waveguide system can in this regime be understood in the context of a waveguide that is chirally coupled to a single emitter. The resonator itself only enhances the coupling between the atom and the waveguide.
In the following, we will introduce this model and briefly discuss the consequences of the chiral light--matter interaction. Furthermore, we show how the atom--resonator--waveguide system can be described in the framework of this picture.

\subsection{Emitter directly coupled to a waveguide}
\label{sec:chiral_wg2}

\begin{figure}
	\centering
	\includegraphics[width=0.6\columnwidth]{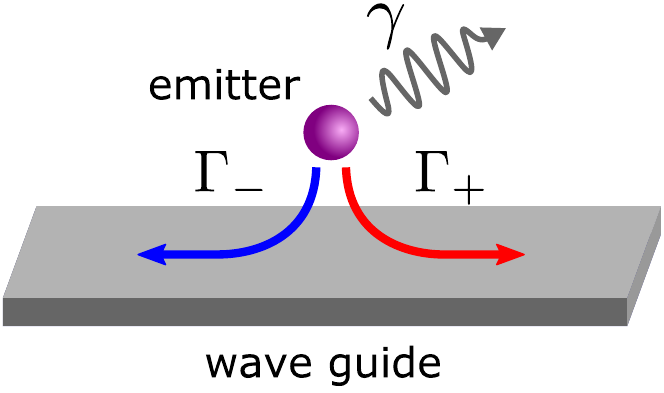}
	\caption{Schematic of a single emitter coupled to a  single mode waveguide. The emitter decays into free space at rate $\gamma$ and couples to the respective forward or backward propagating waveguide mode with $\Gamma_+$ and $\Gamma_-$. }
	\label{fig:Chiral_wave_guide_single}
\end{figure}

\begin{figure*}[tb]
		\includegraphics[width=0.9\textwidth]{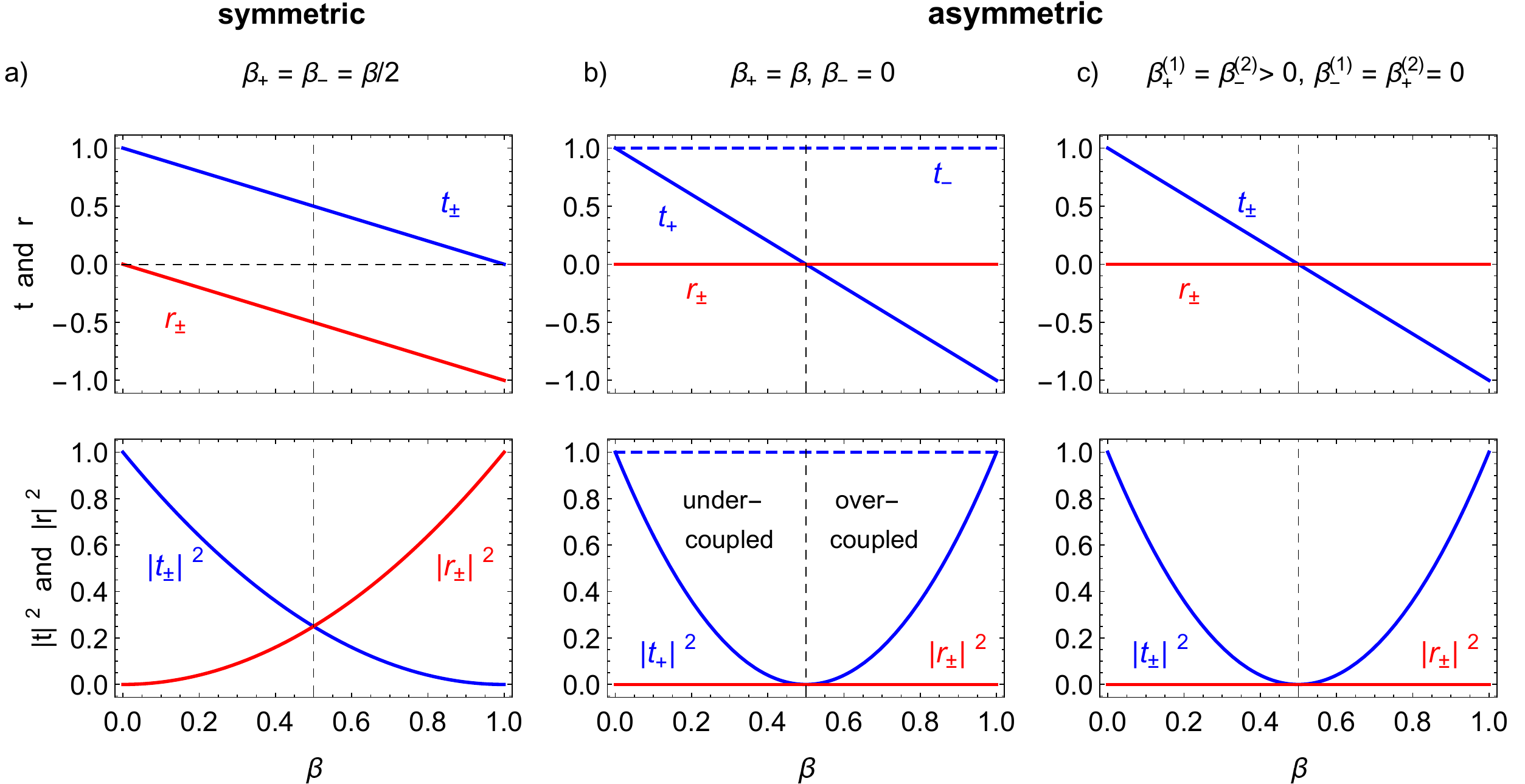}
	\caption{Transmitted and reflected amplitude (top) and power (bottom), normalized to the respective input, as a function of $\beta$. a) Symmetric coupling to the waveguide, i.e. $\beta_+=\beta_-$. For the asymmetric case, we distinguish  b)  $\beta_+=\beta, \beta_-=0$ from  c)  $\beta_+^{(1)}=\beta_-^{(2)}=\beta, \beta_-^{(1)}=\beta_+^{(2)}=0$. In b) the dashed line corresponds to the transmission for the backward probing. }
	\label{fig:plotCouplingRegimes}
\end{figure*}

Let us consider a simple system which consists of a waveguide supporting a pair of counter-propagating modes and a single two-level emitter placed in its vicinity (see Fig.~\ref{fig:Chiral_wave_guide_single}). When the emitter is excited, it can lose its energy either into free space, at a rate $2\gamma$, or into the forward $(+)$ or backward $(-)$ propagating waveguide mode, at rate $2\Gamma_\pm$, respectively. The emission rates into the waveguide modes are governed by $\Gamma_\pm \propto |\boldsymbol{\mu}^* \cdot \boldsymbol{\mathcal{E}}_{\pm}|^2$, where $\boldsymbol{\mathcal{E}}_{\pm}$ is the vector amplitude of the respective mode and $\boldsymbol{\mu}$ is the complex dipole matrix element of the emitter's transition. In the presence of spin--momentum locking, the local field vectors of the two counter-propagating modes can differ, i.e. $\boldsymbol{\mathcal{E}}_{+} \neq \boldsymbol{\mathcal{E}}_{-}$. Thus, the corresponding emission rates are in general not symmetric,  $\Gamma_+\neq \Gamma_-$. For the general case, we can define the two directional coupling parameters
\begin{equation}
\beta_+ =\frac{\Gamma _+}{\Gamma _++\Gamma _-+\gamma} \hspace{0.5cm} \text{and} \hspace{0.5cm} \beta_- =\frac{\Gamma _-}{\Gamma _++\Gamma _-+\gamma} \;,
\end{equation}
which give the ratio of the emission rate into the $(\pm)$ waveguide mode $\Gamma_\pm$ to the total emission rate, including the decay into free space $\gamma$. The two parameters can be combined to give the total coupling efficiency of the emitter into the waveguide, $\beta=\beta_+ +\beta_-$. For many applications, the transmission properties of the waveguide are the most important quantity. The transmission and reflection amplitudes of light propagating in the forward $(+)$ or backward $(-)$ directions in the waveguide, for the case where the light and the emitter are resonant, can be obtained from the directional coupling parameters via
\begin{align}
t_\pm &= 1- 2 \,\beta_\pm\;,\\
r_\pm &= -2\sqrt{\beta_+ \beta_-}\;.
\end{align}
For extending this situation to an emitter with two excited states that couple to orthogonal polarizations we have to define two independent directional coupling parameters $\beta_\pm^{(i)}$, where $i \in \{1,2\}$ indicates the respective transition.
The on-resonance transmission and reflection amplitudes for the three-level system can be obtained from
\begin{align}
t_\pm &= 1- 2 \,(\beta_\pm^{(1)}+\beta_\pm^{(2)})\;,\label{eq:3la1}
\\
r_\pm &= -2\sqrt{\beta_+^{(1)} \beta_-^{(1)}}-2\sqrt{\beta_+^{(2)} \beta_-^{(2)}}\;.
\label{eq:3la2}
\end{align}
Using this simplified model, we can already discuss most of the key features of chiral light--matter interaction.
In Fig.~\ref{fig:plotCouplingRegimes} the transmission (reflection) amplitude and the power transmission (reflection) are plotted as a function of $\beta$ for three important cases. 
In the case of symmetric coupling ($\beta_+\equiv\beta_+^{(1)}+\beta_+^{(2)}=\beta_-\equiv\beta_-^{(1)}+\beta_-^{(2)}$), the emitter equally couples to both probing directions. This situation is realized with a linear emitter or when the waveguide mode does not exhibit SML. The other two cases are realized for fully asymmetric coupling: On the one hand, when the light only couples to a single transition that couples to one probing direction, i.e. $\beta_+>0$ and $\beta_-=0$ and on the other hand, when the emitter has two excited states where each transition exclusively couples to one propagation direction, $\beta_+^{(1)}=\beta_-^{(2)}>0$, $\beta_-^{(1)}=\beta_+^{(2)}=0$.

\begin{figure*}
	\centering
		\includegraphics[width=0.9\textwidth]{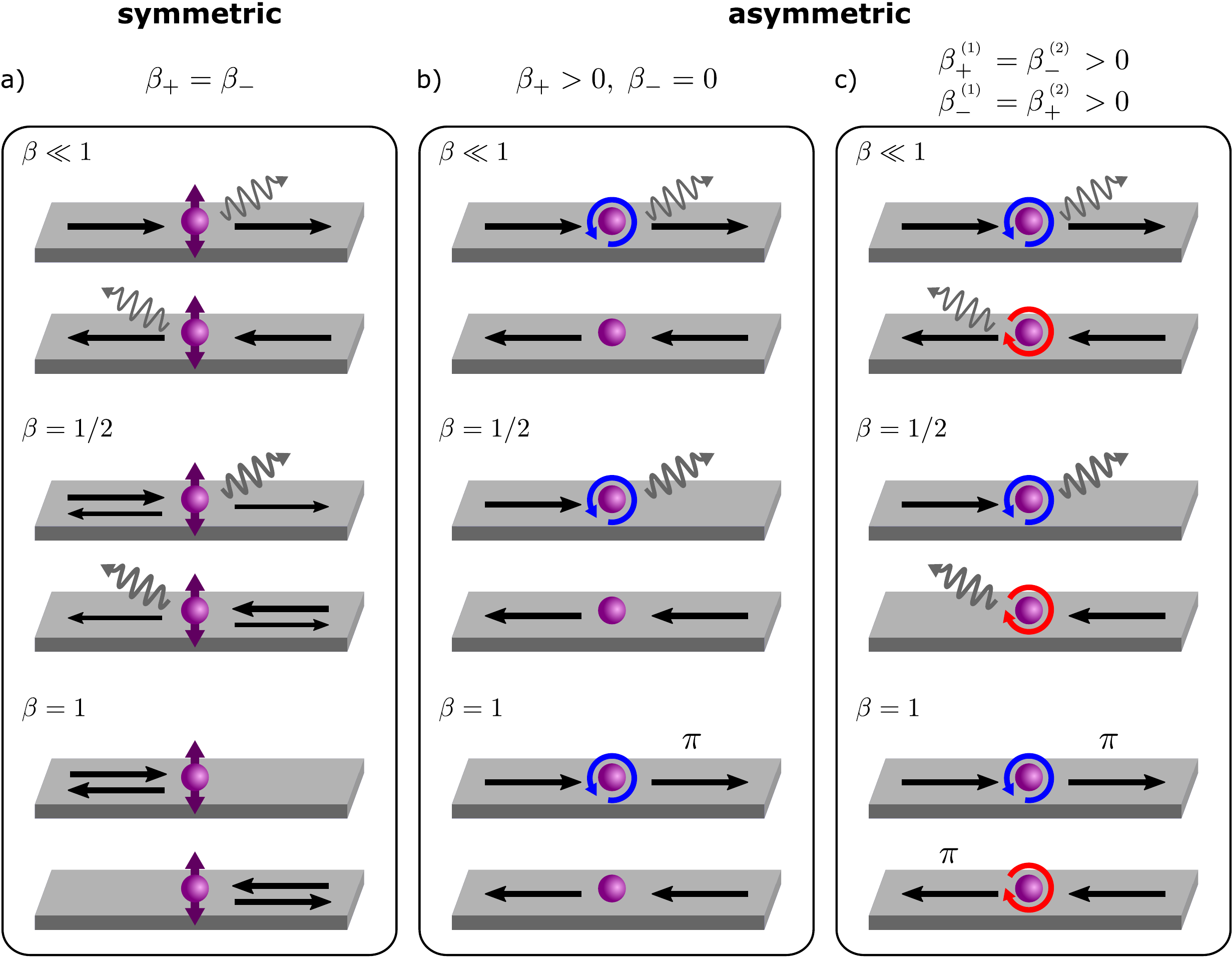}
	\caption{Resonant photon-emitter interaction for symmetric $\beta_+ =\beta_-$ a) and fully asymmetric b) and c) waveguide-emitter coupling. One can distinguish three regimes. $\beta \ll 1$: a) For symmetric coupling, the light in the waveguide is only marginally influenced by the weak coupling to the scatterer. This introduces an additional loss channel to the waveguide, which is independent of the propagation direction. b) If $\beta_+ > 0$, $\beta_-\!=\!0$, only one direction is affected by the presence of the scatterer, giving a small direction-dependent loss. $\beta\!=\!1/2:$ a) The symmetric waveguide reflects and transmits 25\% of the incident light while 50\% is scattered into free space. b,c) For the asymmetric case, all the light is dissipated by the scatterer when it is probed in forward direction. In backward direction, light is either fully transmitted b) or one observes perfect absorption in both directions c). Situation b) can be employed to realize an optical diode. $\beta=1$: a) The scatterer acts as a perfect mirror, reflecting all light incident from one side. b,c) In both directions, the waveguide is perfectly transparent, but the scatterer imprints a phase of $\pi$ onto the forward transmitted light b) or for both propagation directions c). The purple, red and blue arrows indicate the local polarizations of the waveguide modes. For the asymmetric case, the local polarization depends on the propagation direction and the scatterer only couples to the polarization indicated by the arrows. }
	\label{fig:Chiral_wave_guide}
\end{figure*}

\subsubsection{Symmetric coupling}
We will now discuss the case of symmetric coupling, i.e. the case where $\beta_+=\beta_-$. As long as the emitter preferentially couples to non-guided modes ($\beta \ll 1 $) it mainly acts as a loss channel for the guided light (see Fig.~\ref{fig:Chiral_wave_guide}a). As $\beta$ increases, the transmission drops while an increasing fraction of the light is reflected. For $\beta\!=\!0.5$, both transmission and reflection reach 0.25 of the incident power, while half the light will be dissipated into free space.  
When $\beta\!>\!0.5$, the light couples very strongly to the scatterer and is efficiently coupled back into the opposite direction of the waveguide. The transmission steadily decreases and the reflection increases until, for $\beta\!=\!1$, the emitter becomes a perfect mirror, where all the light will be reflected, i.e. $|r_\pm|^2\!=\!1$ and $|t_\pm|^2\!=\!0$, as indicated in Fig.~\ref{fig:Chiral_wave_guide}a. Most importantly,  the transmission and reflection is independent of the probing direction for all $\beta$.
\subsubsection{Fully asymmetric coupling}
For the fully asymmetric case we have to distinguish two situations. In the first case, only one atomic transition couples to the light, i.e. $\beta_-\!=\!0$ and $\beta_+\!>\!0$ and we obtain fully direction-dependent scattering properties of the emitter, meaning that it only interacts with light propagating in one direction. Consequently, the forward propagating light will be modified, i.e. $t_+\!\neq\!1$, while the backward propagating light is not altered and $t_- \!=\! 1$. In addition, the scatterer never reflects the incoming signal, i.e. $r_\pm=0$. As we increase $\beta$, the forward transmission will decrease.   
For $\beta\!=\!0.5$, the coupling rate to free space is equal to the coupling rates into the guided mode, $\Gamma_+ \!=\!\gamma$, and the forward transmission becomes zero. Since the light cannot be reflected, all the light is dissipated by the scatterer, which thus acts as a perfect absorber (see Fig.~\ref{fig:Chiral_wave_guide}d). For $\beta\! >\! 0.5$, the transmission rises again with increasing $\beta$ until the transmission reaches $1$ for  $\beta\!=\!1$. However, for $\beta>0.5$, the interaction with the scatterer imprints an additional phase of $\pi$ onto the transmitted light compared to the uncoupled case.\\
In the second case, we consider an emitter with two excited states and fully direction dependent coupling, i.e. $\beta_+^{(1)}= \beta_-^{(2)}>0$ and $\beta_-^{(1)}= \beta_+^{(2)}=0$ or vice versa. As a consequence, the coupling strength between the light and the emitter is independent of the probing direction and the forward and backward transmission will be the same. However, due to the presence of SML, the light will couple to one or the other atomic transition depending on its propagation direction.

We can summarize the forward transmission properties for the asymmetric case by defining three coupling regimes between the waveguide and the emitter:
\begin{description}
\item [Under-coupling:] {As long as $\beta_+\! <\! 0.5$, the coupling rate to the forward mode is smaller than the coupling rate to all other modes, which include the decay into free space modes and the backward propagating mode, $\Gamma_+ \!<\!\Gamma_- +\gamma$. In this regime, the transmitted light is dominated by the light which did not couple to the scatterer. As $\beta$ approaches $0.5$, the forward transmission continuously decreases because a fraction of the light that interacted with the emitter destructively interferes with the forward propagating light in the waveguide. }
\item [Critical coupling:] {When $\beta_+\!=\!0.5$, the field from the emitter and the field propagating in the waveguide's forward direction have the same amplitude but opposite sign. Thus, they perfectly cancel and the transmission becomes zero. For the symmetric case, critical coupling can only be reached for the limit $\beta\!=\!1$.  }
\item [Over-coupling:] {This regime can only be reached for asymmetric coupling between the waveguide and the emitter. When $\beta_+$ surpasses $0.5$, the light is very efficiently coupled from the waveguide to the scatterer and back into the waveguide. Thus, the amplitude of light that interacted with the scatterer exceeds that of the light that stayed in the waveguide, yielding a finite transmission. However, the interaction with the scatterer introduced an additional phase of $\pi$ compared to the transmitted light in the under-coupled case. }
\end{description}

\subsection{Resonator enhanced chiral waveguide}
\begin{figure*}
	\centering
		\includegraphics[width=1.1\columnwidth]{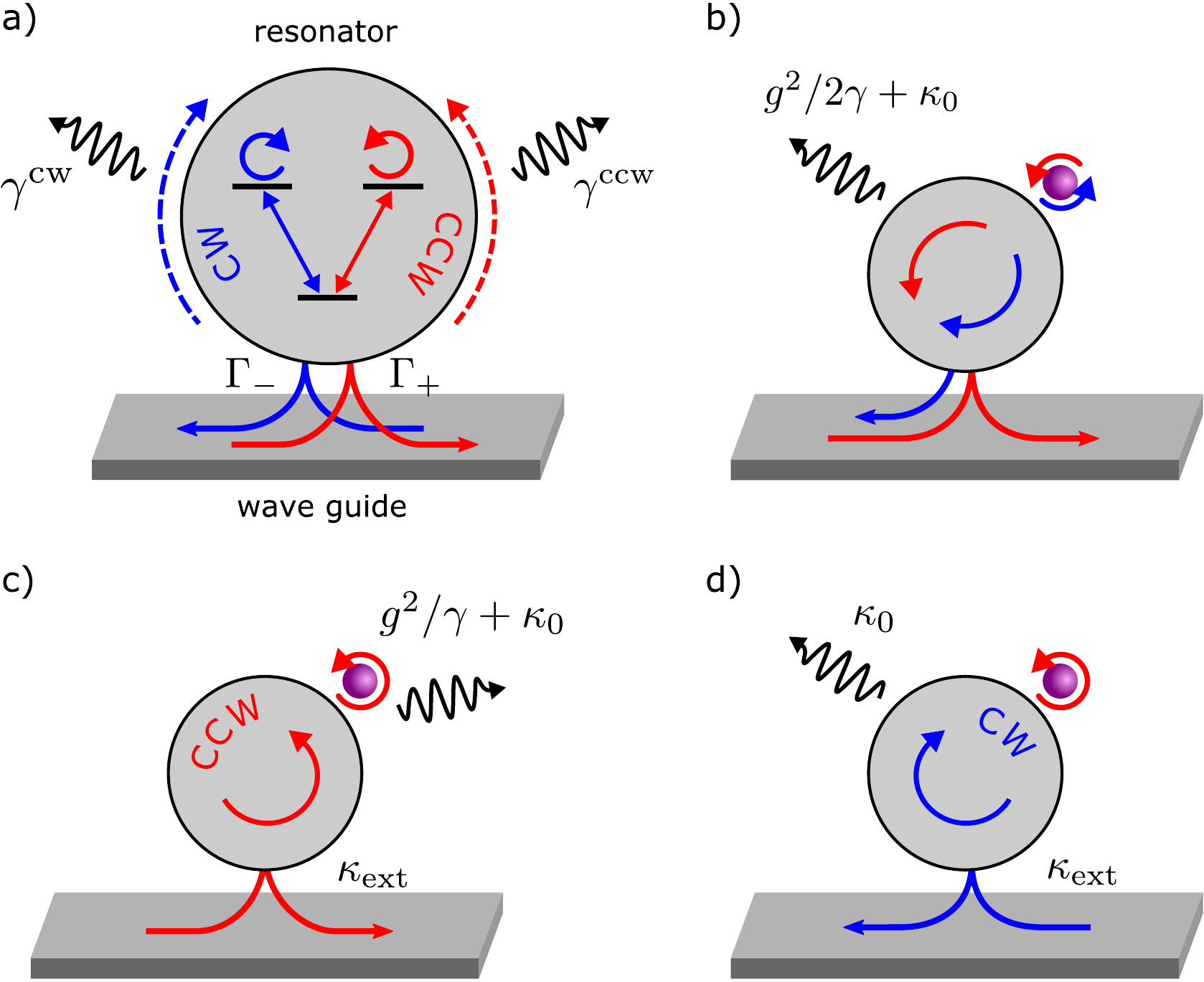}
	\caption{ The coupling between the resonator and the waveguide can be understood as the interaction between the waveguide modes and a $V$-like system. a) For the simple case of an empty resonator, the two excited states correspond to light populating the clockwise (CW) or counter-clockwise (CCW) mode of the resonator, which are coupled to the waveguide modes and are subject to intrinsic losses, $\kappa_0$. In the absence of a scatterer, these modes are not coupled. b) A symmetrically coupled atom, couples the CW and CCW mode, thus giving rise to a finite reflection and transmission. For this system the transmission and reflection properties do not depend on the probing direction. 	
	c) The presence of an atom that only couples to the CCW mode changes the total resonator losses to $g^2/\gamma+\kappa_0$ for light that propagates in the positive direction through the waveguide. d) Since the atom does not couple to the CW mode, the resonator losses and thus the transmission of the light propagating in negative direction is not altered by the atom.  }
	\label{fig:Chiral_wave_guide_resonator}
\end{figure*}

\label{sec:res_enhanced_WG}

Realizing a strong interaction between light guided in a waveguide and a single emitter, i.e. $\beta>0.5$, is experimentally challenging. One way to reach this regime is to make use of the enhanced interaction in optical resonators. In this case, the light in the waveguide is first coupled into a resonator and the resonator field is then coupled to the emitter. \\
The interaction between the waveguide and a WGM resonator, with or without an atom coupled to it, can be described in the framework of chiral waveguides. 
In analogy to the directly coupled emitter considered above, we can define two effective excited states for the atom-resonator system whose coupling to the waveguide is given by $\beta_j^i$, where $i \in \{1,2\}$ and $j \in \{+,-\}$. Similar to the previous section, the transmission and reflection of light in the waveguide is then also covered by eqs (\ref{eq:3la1}) and (\ref{eq:3la1}), respectively.

\subsubsection{Coupled resonator--waveguide system}
\label{sec:perfect chiral}

In order to understand the coupled resonator--waveguide system, it is instructive to first consider the case where no scatterer is coupled to the resonator modes. The resonator can then be interpreted as a three-level scatterer with two excited states, where the excited states correspond to having a photon in the CW or CCW mode, as illustrated in Fig.~\ref{fig:Chiral_wave_guide_resonator}a. 
For the empty resonator, each of the two transitions only couples to one of the two propagation directions of the waveguide mode. If the resonator has been excited from one propagation direction, it can only decay into the waveguide mode that propagates in the same direction. Thus, the empty resonator represents a fully chiral system, see Fig. \ref{fig:Chiral_wave_guide}c. The corresponding $\beta$-factors are
\begin{align}
\beta_+^\text{ccw}|_{_{g=0}}&= \beta_-^\text{cw}|_{_{g=0}}=\frac{\kappa_\text{ext} }{\kappa_0 + \kappa_\text{ext}}\;,\\
\beta_-^\text{ccw}|_{_{g=0}}&=\beta_+^\text{cw}|_{_{g=0}}=0\;.
\end{align} 
yielding the direction independent on-resonance transmission amplitude 
\begin{equation}
t_\pm|_{_{g=0}}=\frac{\kappa_0-\kappa_\text{ext}}{\kappa_0+\kappa_\text{ext}}\;.
\label{eq:emptyresonator}
\end{equation}
From this equation it also becomes evident that we are able to over-couple the empty resonator if $\kappa_\text{ext}>\kappa_0$. In analogy to the situation depicted in Fig. \ref{fig:Chiral_wave_guide}c, the transmission past the empty resonator is independent of the probing direction and, since the two modes of the resonator do not couple, the reflection amplitude is always zero, $r_\pm=0$. \\
Again, we can assign three different coupling regimes:
The under-coupled regime, for which the internal resonator losses are larger than the resonator--waveguide coupling. If $\kappa_\text{ext}=\kappa_0$, the transmission drops to zero, which is called critical coupling. And, finally, when the coupling to the waveguide becomes the dominate rate, i.e. $\kappa_\text{ext}>\kappa_0$, we speak of an over-coupled resonator.
\subsubsection{Coupled two-level atom--resonator--waveguide system}
\begin{figure}[tbh]
	\centering
		\includegraphics[width=0.9\columnwidth]{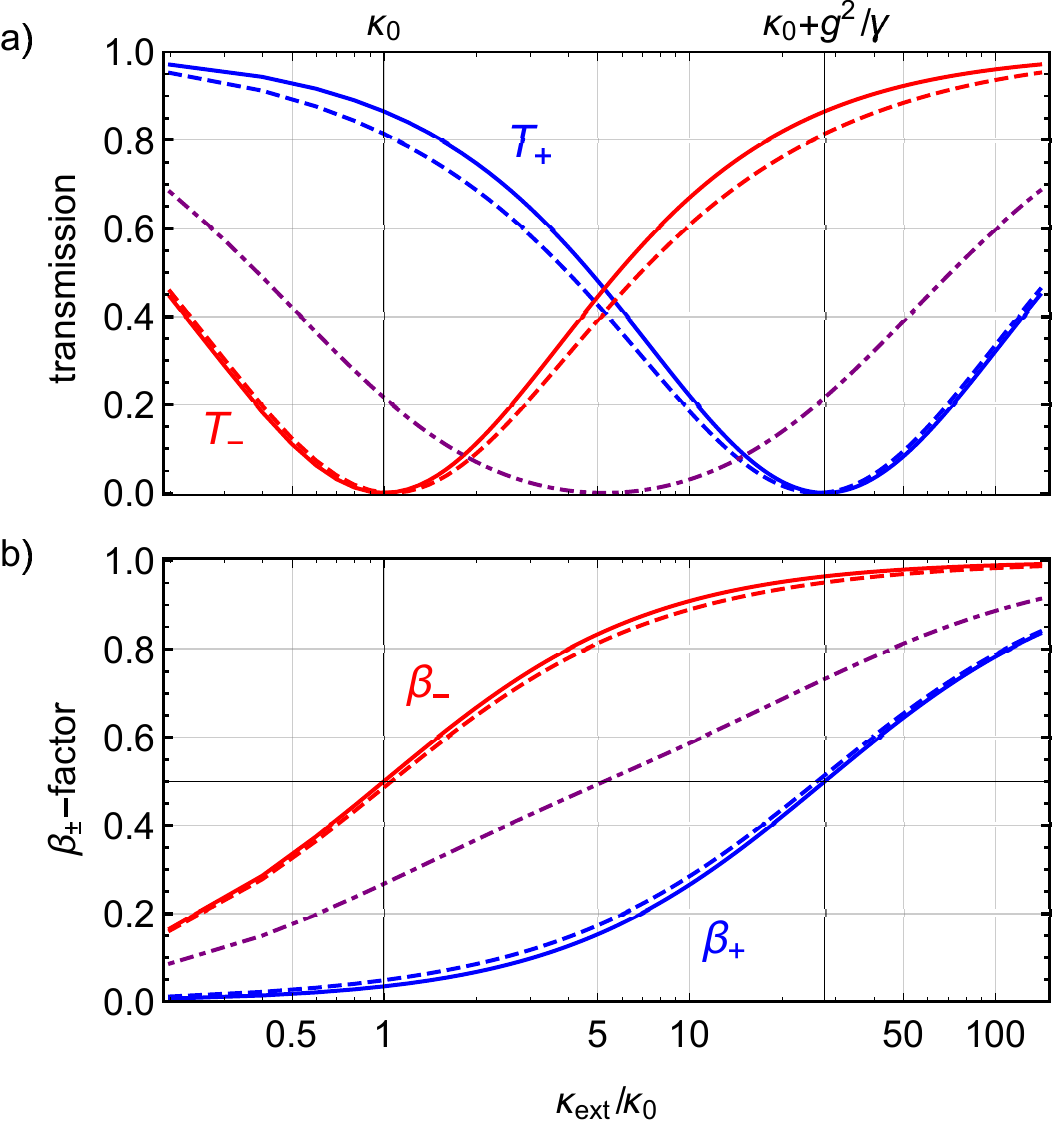}[tb]
	\caption{a) On-resonant forward, T$_+\!=\!|t_+|^2$, and backward, T$_-\!=\!|t_-|^2$, transmission through the waveguide coupled to a resonator--atom system as a function of the coupling rate between the resonator and the waveguide, in units of the intrinsic resonator loss rate $\kappa_0$. The transmission is plotted for three different polarization settings: perfect chiral coupling, i.e. $|\alpha_{\sigma^+}|^2\!=\!1$ (solid lines), the polarization obtained for silica WGMs, i.e. $|\alpha_{\sigma^+}|^2\!=\!0.97$ (dashed lines) and for the non-chiral case $|\alpha_{\sigma^+}|^2\!=\!0.5$ (dash-dotted line). 
Note that $T_-$ for $|\alpha_{\sigma^+}|^2\!=\!1$ corresponds to the transmission in the absence of the atom.
b) Total $\beta$-factor for forward and backward probing, i.e. $\beta_\pm=\beta_\pm^\text{ccw}+\beta_\pm^\text{cw}$, as a function of $\kappa_\text{ext}$. Critically coupling, i.e. $T_\pm=0$, requires $\beta_\pm\!=\!0.5$. In both a) and b) the following typical parameters were used: $(g,\gamma,\kappa_0)=2\pi\times(20,3,5)$~MHz.    
	}
	\label{fig:plotbeat_BMR}
\end{figure}

Now we consider the case wherean emitter is coupled to the resonator. For sake of simplicity, we will restrict our further discussion to the most interesting case of emitters that exhibit a single excited state that is coupled via a $\sigma^+$-polarized transition. The polarization state of the resonator modes is determined by the polarization overlap $|\alpha_{\sigma^+}|^2$. For this situation the $\beta$-factors are given by
\begin{align}
\beta_+&=\frac{ \kappa_\text{ext}}{\kappa_0+\kappa_\text{ext}}\frac{g^2(1-|\alpha_{\sigma^+}|^2)/\gamma+\kappa_0+\kappa_\text{ext}}{g^2/\gamma+\kappa_0+\kappa_\text{ext}} \;,\\
\beta_-&=\frac{ \kappa_\text{ext}}{\kappa_0+\kappa_\text{ext}}\frac{g^2|\alpha_{\sigma^+}|^2/\gamma+\kappa_0+\kappa_\text{ext}}{g^2/\gamma+\kappa_0+\kappa_\text{ext}} \;.
\end{align}
The corresponding on-resonant transmission amplitudes are
\begin{align}
t_+ &=\frac{  \left(\kappa _0^2-\kappa _{\text{ext}}^2\right)+g^2 \left((2 |\alpha_{\sigma^+}|^2-1 )\kappa _{\text{ext}}+\kappa
   _0\right)/\gamma}{\left(\kappa _{\text{ext}}+\kappa _0\right) \left(  \left(\kappa _{\text{ext}}+\kappa _0\right)+g^2/\gamma\right)}\;,   \label{eq:twg1}\\
t_- &=\frac{  \left(\kappa _0^2-\kappa _{\text{ext}}^2\right)+g^2 \left((1-2 |\alpha_{\sigma^+}|^2 )\kappa _{\text{ext}}+\kappa
   _0\right)/\gamma}{\left(\kappa _{\text{ext}}+\kappa _0\right) \left(  \left(\kappa _{\text{ext}}+\kappa _0\right)+g^2/\gamma\right)}\;. \label{eq:twg2}
\end{align}
The on-resonance transmission through a waveguide that is coupled to an atom--resonator system as a function of $\kappa_\text{ext}$ is shown for different polarization settings $|\alpha_{\sigma^+}|^2$ in Fig.~\ref{fig:plotbeat_BMR}a. Note that due to SML the forward and backward transmission are related via $t_-(|\alpha_{_{\sigma^+}}|^2)\!=\!t_+(1-|\alpha_{_{\sigma^+}}|^2)$.
For $|\alpha_{\sigma^+}|^2$ close to one we get the chiral behavior described in Sec.~\ref{sec:perfect_chiral}, i.e. the transmission is significantly different for the two probing directions. 

In addition, Fig.~\ref{fig:plotbeat_BMR}b shows $\beta_\pm$ as a function of $\kappa_\text{ext}$. Note that, whenever $\beta_\pm$ is $0.5$, the respective transmission becomes zero. The difference between $\beta_-$ and $\beta_+$ is maximal for $\kappa_\text{ext}= \sqrt{\kappa_0(g^2/\gamma+\kappa_0)}$. Interestingly, for this coupling the forward and backward transmission become equal.

\paragraph{Symmetric atom--resonator system}\mbox{}\\
\label{sec:symmetric_coupling}
Let us now consider the special case of symmetric coupling. This situation can be realized by either having linearly polarized resonator modes that do not exhibit SML, i.e $|\alpha_{\sigma^+}|^2\!=\!|\beta_{\sigma^-}|^2\!=\!1/2$, or by coupling a linearly polarized two-level emitter to the resonator.
In both cases the emitter couples equally to both counter-propagating modes. As a consequence, we obtain the same loss rates for the two counter-propagating resonator modes $\kappa_0+g^2/2\gamma$ and from eqns (\ref{eq:twg1}) and(\ref{eq:twg2}) we obtain for the $\beta$ factors of the symmetric system  
\begin{align}
\beta_\pm|_{_{|\alpha|^2=1/2}}&=\frac{ \kappa_\text{ext}}{\kappa_0+\kappa_\text{ext}}\frac{g^2/2\gamma+\kappa_0+\kappa_\text{ext}}{g^2/\gamma+\kappa_0+\kappa_\text{ext}} \;.\\
\end{align}
The on-resonant transmission and reflexion amplitudes are
\begin{align}
 t_\pm|_{_{|\alpha|^2=1/2}}&=\frac{\kappa_0+\frac{g^2}{\gamma}\frac{\kappa_0}{\kappa_0+\kappa_\text{ext}}-\kappa_\text{ext}}{\kappa_0+g^2/\gamma+\kappa_\text{ext}}\\
r_\pm|_{_{|\alpha|^2=1/2}}&=\frac{\frac{g^2}{\gamma}\frac{\kappa_\text{ext}}{\kappa_0+\kappa_\text{ext}}}{\kappa_0+g^2/\gamma+\kappa_\text{ext}}\;.
\end{align}
The transmission becomes zero for $\kappa_\text{ext}= \sqrt{\kappa_0(g^2/\gamma+\kappa_0)}$. As we will see later, this corresponds to the geometrical mean of the forward and backward critical coupling points of the chiral waveguide.\\
\paragraph{Asymmetric atom--resonator system}\mbox{}\\
\label{sec:perfect_chiral}
Let us now consider the special case where the resonator modes exhibit perfect circular polarization, i.e $|\alpha_{\sigma^+}|^2\!=\!|\beta_{\sigma^-}|^2\!=\!1$. Thus, an atom that exhibits a $\sigma^+$-transition exclusively couples to the CW mode. As a consequence, we obtain two different loss rates $\kappa_0$ and $\kappa_0+g^2/\gamma$ for the CW and the CCW resonator mode,respectively. 
Here, the CW mode has the intrinsic resonator loss rate, while the CCW mode is in addition subject to an atom-induced loss rate $g^2/\gamma$.
At the same time, the waveguide coupling rate for the two probing directions is given by $\kappa_\text{ext}$.
This gives rise to the following $\beta$-factors 
\begin{align}
\beta_+|_{_{|\alpha|^2=1}}&=\frac{ \kappa_\text{ext}}{g^2/\gamma+\kappa_0+\kappa_\text{ext}} \;,\\
\beta_-|_{_{|\alpha|^2=1}}&=\frac{\kappa_\text{ext} }{\kappa_0 + \kappa_\text{ext}}\;.
\end{align}
The corresponding on-resonant transmission amplitudes in forward and backward direction are
\begin{align}
 t_+|_{_{|\alpha|^2=1}}&=\frac{\kappa_0+g^2/\gamma-\kappa_\text{ext}}{\kappa_0+g^2/\gamma+\kappa_\text{ext}}\;,\\
 t_-|_{_{|\alpha|^2=1}}&=\frac{\kappa_0-\kappa_\text{ext}}{\kappa_0+\kappa_\text{ext}}\;.
\end{align}
For the backward probing direction we recover the empty resonator. In forward direction, the presence of the atom introduces additional losses, that shift the critical coupling point to $\kappa_\text{ext}\!=\!\kappa_0+g^2/\gamma$. Thus, the interaction with the atom, can place a formally critically or over-coupled resonator into a different coupling regime. As the atom only interacts with one of the two counter-propagating resonator modes, this yields a direction-dependent, i.e. chiral, transmission. 
Note that these transmission properties are different to those discussed in the previous section. There a single emitter is coupled to a chiral waveguide and the direction-dependent waveguide--emitter coupling stems from the unbalanced emission rate into the two waveguide dirations. In the situation discussed in this section, the direction-dependence stems from the modified resonator loss rates which results in direction-dependent losses.\\

If we consider two excited states with orthogonal polarization, the light will couple to the emitter independent of the probing direction. Thus, the forward and backward transmission will be the same. The corresponding direction-independent $\beta$-factors are 
\begin{align}
\beta_+|_{_{|\alpha|^2=1}}&=\beta_-|_{_{|\alpha|^2=1}}=\frac{ \kappa_\text{ext}}{g^2/\gamma+\kappa_0+\kappa_\text{ext}} \;,
\end{align}
and give rise to the transmission amplitudes
\begin{align}
 t_+|_{_{|\alpha|^2=1}}&= t_-|_{_{|\alpha|^2=1}}=\frac{\kappa_0+g^2/\gamma-\kappa_\text{ext}}{\kappa_0+g^2/\gamma+\kappa_\text{ext}}\;. \label{eq:tdiode}
\end{align}

\section{Applications of chiral light--matter interaction}\label{chap:diode}
In the following, we will briefly discuss three applications that make use of chiral light--matter interaction in WGM resonators. The first two employ a polarization depending scatterer to break Lorentz reciprocity, realizing integrated optical isolators and circulators. The last example uses a three-level $\Lambda$-system to realize a bistable system that changes its properties after each photon scattering. This is the base for a deterministically operating passive quantum memory and a universal photon--photon quantum gate.
\subsection{Optical diode}

\begin{figure}
	\centering
		\includegraphics[width=1.0\columnwidth]{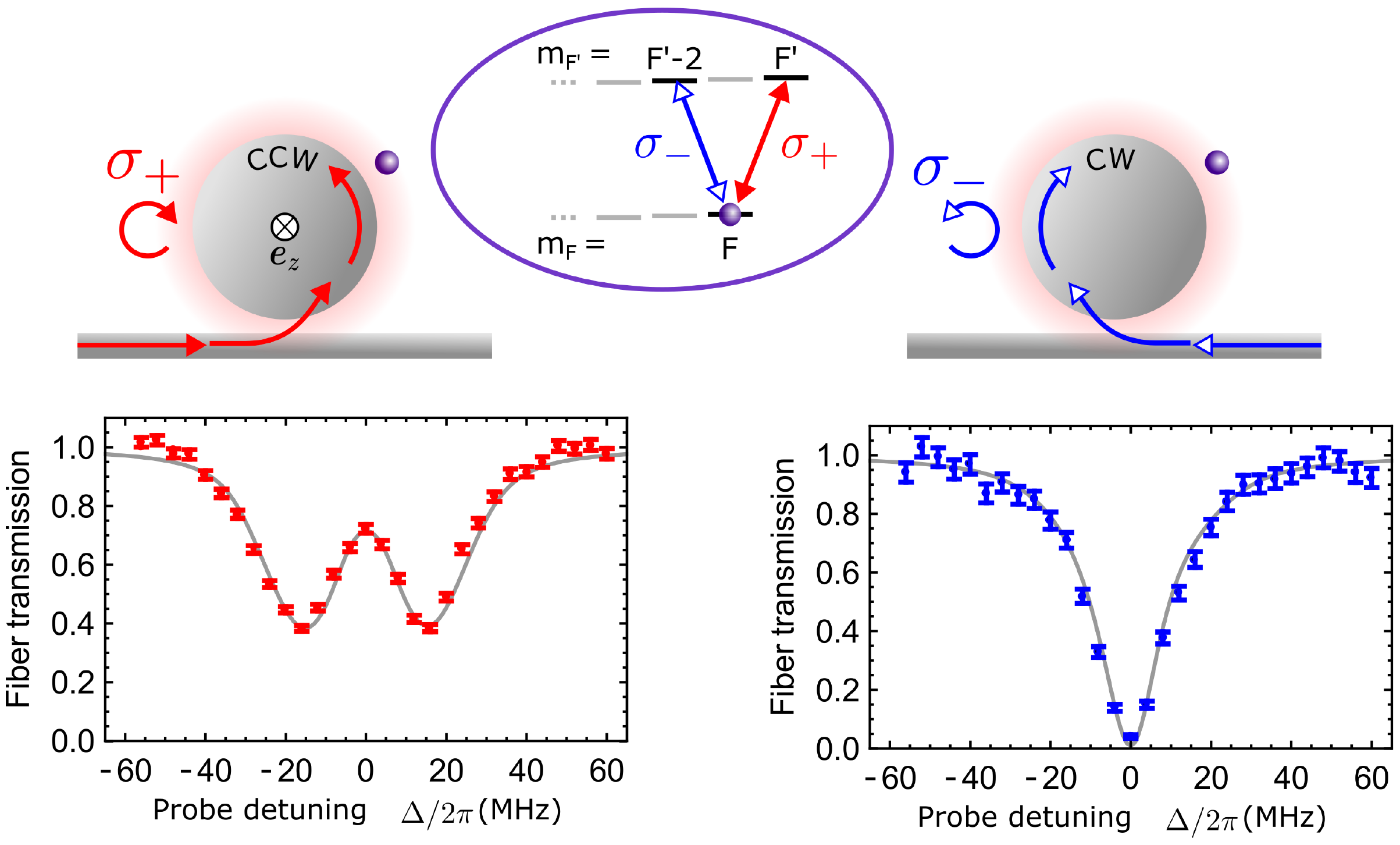}
	\caption{Working principle of an optical diode. The atom is prepared in a state such that it predominately couples to light in the CCW mode. The resonator is interfaced with one coupling waveguide. The transmission spectra are taken from ref \cite{Junge2013} where the system is probed from the left (red data)  and from the right (blue data). The solid lines in the spectra corresponds to a theoretical predictions taking into account the full level structure and correct polarization of the resonator modes and a coupling strength distribution. On-resonance light is only transmitted in one direction. \cite{Sayrin2015bb} }
	\label{fig:diode_resonator_atom}
\end{figure}
To realize an optical isolator by making use of SML and a polarization-dependent scatterer was proposed in Ref.~\cite{Lenferink2014}.
For this scheme, the resonator is interfaced by a single waveguide (see Fig.~\ref{fig:diode_resonator_atom}a). The two ends of the coupling waveguide are referred to as port 1 and port 2, respectively. 
We now consider the case where light that propagates in CW direction in the resonator is almost fully $\sigma^-$-polarized, while it is almost fully $\sigma^+$-polarized when it propagates in the CCW direction. These two modes are driven via the respective probing directions through the coupling waveguides. A single atom is prepared in the outermost Zeeman sublevel $m_F\!=\!+F$ of the hyperfine ground state. Thus, the resonator fields then drive an effective V-system. The strength of one transition is much stronger than the competing transition. Consequently, light that couples from one of the waveguides into the CCW resonator mode couples strongly to the atom with the coupling constant $g_\text{ccw}$. In contrast, light coming from the opposite direction and thus coupling to the CW mode shows only negligible coupling $g_\text{cw}\ll g_\text{ccw}$. The presence of the atom modifies the power transmission spectrum, which in the approximation of perfectly circularly polarized modes is given by

\begin{align}
T_{12(21)}(\Delta)&=\left|\frac{\Gamma_\text{ccw(cw)}+\kappa_0-\kappa_\text{ext}+i\Delta}{\Gamma_\text{ccw(cw)}+\kappa_0+\kappa_\text{ext}+i\Delta}\right|^2\;.  
\label{eq:atom_res_diode}
\end{align}
Here, the subscript indicates the involved input and output ports, respectively and $\Delta$ is the detuning between the resonator -- resonant which the atomic transition -- and the probing light.
In Eq.~(\ref{eq:atom_res_diode}), we have used 
\begin{equation} 
\Gamma_\text{cw/ccw}=\frac{g^2_\text{cw/ccw}}{\gamma+i\Delta}\;,
\label{eq:atom_loss_rate}
\end{equation}
for the additional resonator loss rate introduced by the atom. The presence of the atom changes the total resonator loss rate from $\kappa_\text{tot}\equiv\kappa_0 +\kappa_\text{ext}$ to $\kappa_\text{tot}+\Gamma_\text{cw/ccw}$. For the case where the light couples to the CW resonator mode, $g_\text{cw}$ and thus the atom-induced loss rate is small, i.e. $\kappa_\text{tot}\gg \Gamma_\text{cw}$, and the resonator transmission is not significantly modified by the atom. for a critically coupled resonator, no light will be transmitted on resonance. However, for the CCW direction, $\Gamma_\text{ccw}$ can become comparable to or larger than $\kappa_\text{tot}$. In this case, the resonator--atom system is now in the under-coupled regime and the incident light field will be transmitted through the waveguide. The direction-dependent coupling thus breaks Lorentz reciprocity and realizes an optical diode.
\\
In the experiment described in Ref. \cite{Sayrin2015bb}, the resonator--waveguide coupling rate was adjusted such that $\kappa_0=\kappa_\text{ext}$, which corresponds to critical coupling to the empty resonator. 
A single $^{85}$Rb atom was prepared in the outermost Zeeman sublevel $m_F\!=\!+3$ of the $5S^2_{1/2}$, $F\!=\!3$  hyperfine state.
For this setting, the strength of the transition to the $5P^2_{3/2}$, $\ket{F'\!=\!4,m_{F'}\!=\!+4}$ excited state is 28 times stronger than the one of the transition to the $\ket{F'\!=\!4,m_{F'}\!=\!+2}$ state.
When an atoms is present in the resonator field and the system is probed from one side, a splitting of the spectrum is observed (see Fig.~\ref{fig:diode_resonator_atom}b). As a consequence the on-resonance transmission significantly increases. However, when the system is probed from the opposite direction, the atom is only weakly coupled, thus the spectrum resembles that of an empty resonator (see Fig.~\ref{fig:diode_resonator_atom}c). Thus, on resonance the light is blocked. For this setting, an isolation of  $\mathcal{I}=10\, \text{log}(T_{12}/T_{21})=13$~dB could be realized, only limited by experimental imperfections. Interestingly, preparing the atom in the opposite Zeeman ground state $\ket{F\!=\!3,m_F\!=\!-3}$ changes the role of the CW and CCW mode and yields an isolator with reversed operation direction.

\subsection{Quantum circulator}

\begin{figure}
	\centering
		\includegraphics[width=\columnwidth]{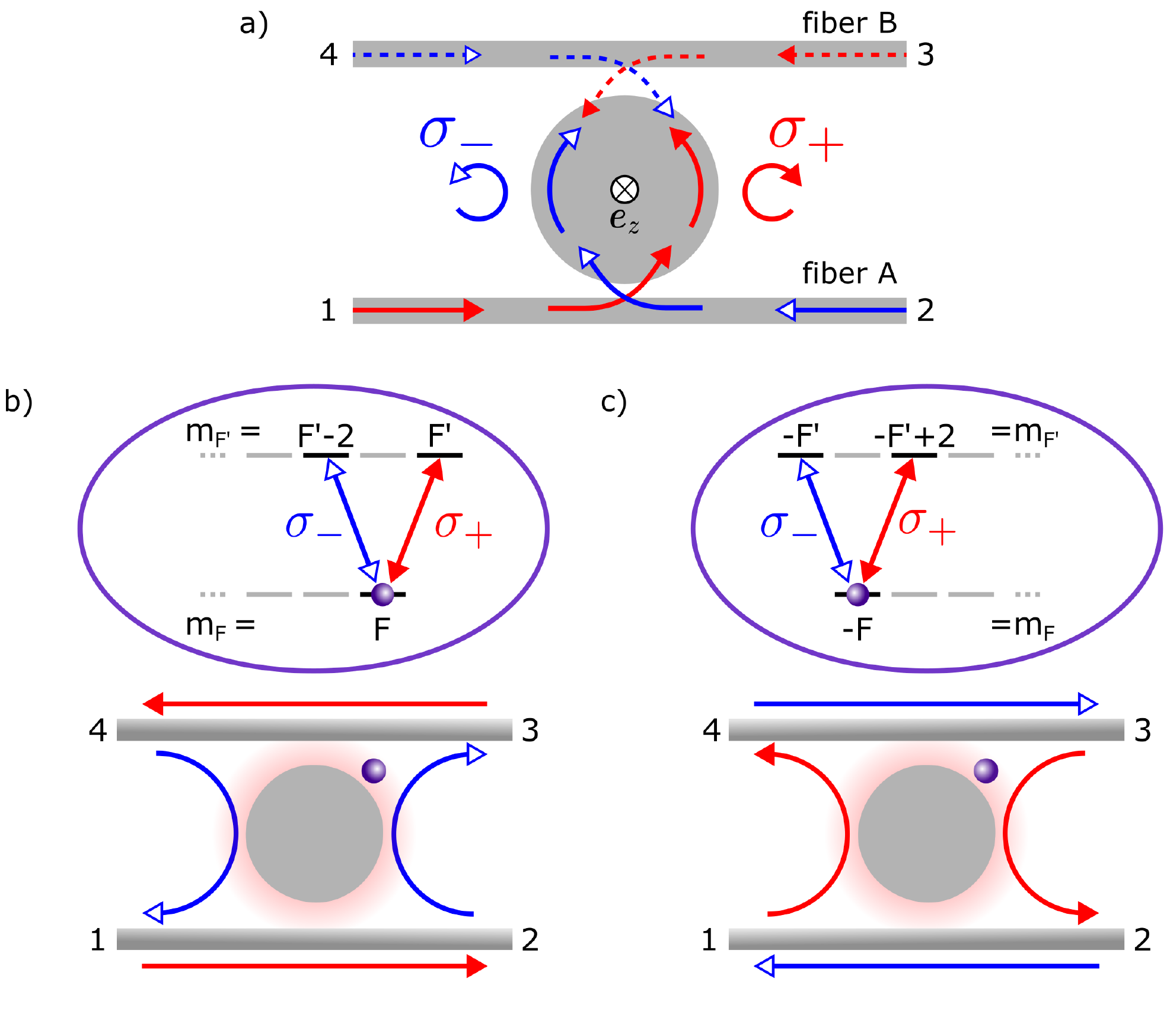}
	\caption{Working principle of an optical circulator. a) Schematic of a resonator interfaced with two coupling waveguides, labeled $A$ and $B$. When probing from port~1 or 3 light is coupled into the CCW propagating resonator mode $a$, for which the local polarization of the evanescent field is $\sigma^+$. From port~2 and 4 the CW propagating $b$ mode is excited, which exhibits $\sigma^-$ polarization. b) \& c) Working principle of the programmable circulator. b) When the atom is prepared in the $m_F\!=\!+3$ hyperfine ground state the circulator operation is defined as $(1\rightarrow 2\rightarrow 3\rightarrow 4\rightarrow1)$. c) When the atom is prepared in the $m_F\!=\!-3$ hyperfine ground state the circulator operation is reversed and now defined as $(1\rightarrow 4\rightarrow 3\rightarrow2\rightarrow1)$.   }
	\label{fig:circulator_resonator_atom}
\end{figure}
The concept of the optical circulator is a straight forward extension of the optical isolator \cite{Xia2014}. The key idea is to replace the dissipative channel in the backward direction with an accessible output. This can be realized by interfacing the WGM resonator with two coupling waveguides, thus forming a 4-port device, as shown in Fig.~\ref{fig:circulator_resonator_atom}a. \\
Let us first consider the case of an empty resonator. From the two coupling waveguides, which are denoted $A$ and $B$, light couples into and out of the resonator at a rate given by $\kappa_A$ and $\kappa_B$ respectively. Thus, the on-resonance power transmission through  waveguide $A$ is given by 
\vspace*{0.5cm}
\begin{align}
T^0_{12}=T^0_{21}&=\left|\frac{\kappa_0+\kappa_{B}-\kappa_{A}}{\kappa_0+\kappa_A+\kappa_B}\right|^2\;, 
\label{eq:empty_res_circ1}
\end{align}
and through waveguide $B$ 
\begin{align}T^0_{34}=T^0_{43}&=\left|\frac{\kappa_0+\kappa_{A}-\kappa_{B}}{\kappa_0+\kappa_A+\kappa_B}\right|^2 \;.
\label{eq:empty_res_circ2}
\end{align}
Here, the subscript indicates the involved ports. The transmission from one waveguide into the other via the resonator is given by  
\begin{align}
T^0_{14}=T^0_{23}=T^0_{32}=T^0_{41}&=\frac{4\, \kappa_A\,\kappa_B}{|\kappa_0+\kappa_A+\kappa_B|^2}\;.
\label{eq:empty_res_circ3}
\end{align}
From Eq.~(\ref{eq:empty_res_circ1})-(\ref{eq:empty_res_circ3}) we see that the empty resonator is reciprocal, since $T_{i,j}^0=T_{j,i}^0$.
In order to achieve efficient, low-loss photon routing, let us assume that the waveguide--resonator coupling rates $\kappa_A$ and $\kappa_B$ are adjusted such that both waveguides are almost critically coupled to the resonator, i.e. $\kappa_A\approx\kappa_B\gg\kappa_0$ . In this setting, all the light sent from one waveguide will be coupled into the resonator and transferred to the opposite waveguide.\\
In order to achieve nonreciprocal light transmission, we again make use of the inherent link between propagation direction and polarization of the resonator fields and couple them to a polarization-dependent scatterer. The presence of the perfectly chirally coupled scatterer modifies the on-resonance power transmission, which is given by
\begin{align}
T_{12(21)}&=\left|\frac{\Gamma_\text{ccw(cw)}+\kappa_0+\kappa_{B}-\kappa_{A}}{\Gamma_\text{ccw(cw)}+\kappa_0+\kappa_A+\kappa_B}\right|^2\;, \label{eq:empty_res_circ1b} \\
T_{34(43)}&=\left|\frac{\Gamma_\text{ccw(cw)}+\kappa_0+\kappa_{A}-\kappa_{B}}{\Gamma_\text{ccw(cw)}+\kappa_0+\kappa_B+\kappa_A}\right|^2\;, \label{eq:empty_res_circ2b}\\ 
T_{14(41)}=T_{32(23)}&=\frac{4 \kappa_A\,\kappa_B}{|\Gamma_\text{ccw(cw)}+\kappa_0+\kappa_A+\kappa_B|^2}\;,
\label{eq:empty_res_circ3b}
\end{align}
where the additional loss rate introduced by the atom, $\Gamma_\text{cw/ccw}$, is defined in Eq.~(\ref{eq:atom_loss_rate}). 
Since $\Gamma_{\text{ccw}}\neq\Gamma_{\text{cw}}$, Lorentz reciprocity is again broken. To realize a circulator, the transmissions in Eqs.~(\ref{eq:empty_res_circ1b}) - (\ref{eq:empty_res_circ3b}) have to be chosen such that one realizes the routing operation defined as $(1\rightarrow 2\rightarrow 3\rightarrow 4\rightarrow1)$. Thus, we have to find a trade-off between efficient light transfer from one waveguide to the other via the weakly coupled CW mode, which implies $(\kappa_A,\kappa_B )\gg \kappa_0+\Gamma_\text{cw}$, and the condition that the presence of the atom should significantly influence the field decay rate, $\Gamma_\text{ccw} \ll \kappa_0+\kappa_A+\kappa_B$. 

An experimental realization of such an optical circulator is described in Ref.~\cite{Scheucher2016}. Here, a single $^{85}$Rb atom that is prepared in the $m_F\!=\!+3$ of the $5S^2_{1/2}$, $F\!=\!3$ hyperfine state is coupled to a WGM resonator and controls the circulator direction.
Preparing the atom in the opposite Zeeman ground state, $\ket{F\!=\!3,m_F\!=\!-3}$, changes the role of the CW and CCW mode and yields a circulator with reversed operation direction, as shown in Fig.~\ref{fig:circulator_resonator_atom}c. This enables one to reconfigure the operation direction of the circulator using the internal spin-state of a single atom. At the optimal working point, $\kappa_\text{tot}/2\kappa_0=2.2$, an operation fidelity exceeding 0.7 was demonstrated\cite{Scheucher2016}. In addition to the high isolation, the device exhibits a low insertion loss of $L=1.4$~dB. \\

\subsection{Single-photon Raman interaction}

\begin{figure}
	\centering
		\includegraphics[width=0.9\columnwidth]{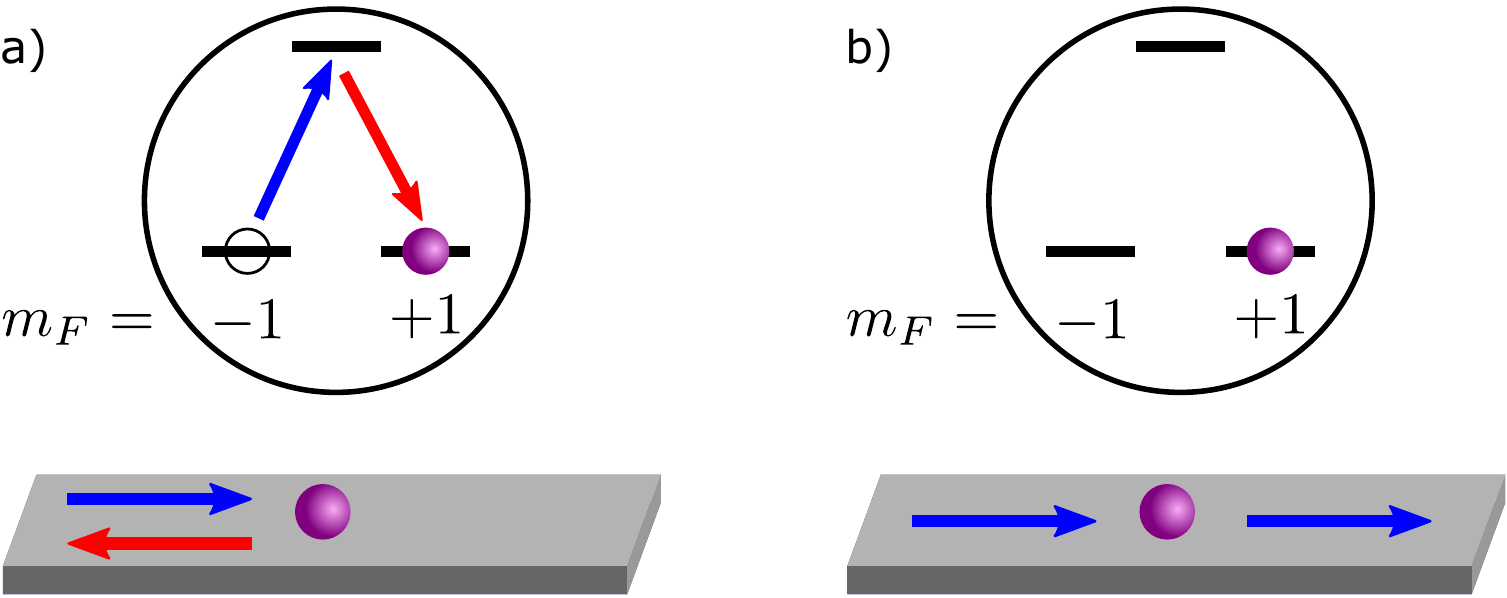}
	\caption{ An atom exhibiting a three-level $\Lambda$-configuration is coupled to a chiral waveguide. As a consequence, the $\sigma_+$ transition is coupled only to the mode propagating to the right, and the $\sigma_-$ transition is coupled only to the mode propagating to the left. a) For an atom in the $m_F=+1$ ground state, the incoming $\sigma_+$ photon will be deterministically reflected as a $\sigma_-$ photon. Thereby, the atomic state is transferred to $m_F =+1$. b) An atom in the $m_F =+1$ ground state does not interact with $\sigma_+$ photons, which are accordingly transmitted. Note that the entire scheme is left-right symmetric \cite{Shomroni2014}.
	}
	\label{fig:swap}
\end{figure}

Another application of light--matter interaction in chiral waveguides is to use a three-level atom that forms a $\Lambda$-system, as shown in Fig.~\ref{fig:swap}. For a perfectly chiral waveguide, light propagating in $\pm$-direction  couples to different atomic transitions. If the two ground-states and the excited state are associated with different total angular momentum. The CCW mode interacts only with the $\sigma^+$-transition between the ground state $\ket{m_F=-1}$  and the excited state $\ket{e}$, whereas the CW mode interacts only with the $\sigma^-$-transition associated with the ground state $\ket{m_F=+1}$.
For equal transition strengths, depending on the probing direction ($\pm$) and the initial ground state ($\ket{m_F=\mp1}$ ) the quasi-steady state transmission and reflection amplitudes are given by \cite{Rosenblum2017}
\begin{align}
t_+^{-1}=t_-^{+1}&=\frac{\kappa_0+\frac{2g^2}{\gamma}\frac{\kappa_0}{\kappa_0+\kappa_\text{ext}}-\kappa_\text{ext}}{\kappa_0+g^2/\gamma+\kappa_\text{ext}}\\
t_-^{-1}=t_+^{+1}&=\frac{\kappa_0-\kappa_\text{ext}}{\kappa_0+\kappa_\text{ext}}\\
r_+^{-1}=r_-^{+1}&=\frac{\frac{2g^2}{\gamma}\frac{\kappa_\text{ext}}{\kappa_0+\kappa_\text{ext}}}{\kappa_0+g^2/\gamma+\kappa_\text{ext}}\\
r_-^{-1}=r_+^{+1}&=0\,.
\end{align}
By choosing $\kappa_\text{ext}=\sqrt{\kappa_0(\kappa_0+2g^2/\gamma)}$, one reaches the critically coupled situation. In this case, when the atom is prepared in the state $|m_f=-1\rangle$ ($|m_f=+1\rangle$) a single incident photon incident from the left (right) will deterministically transfer the atom to the other ground state which is accompanied by the emission of a photon in the opposite direction. The result is a single-photon Raman transfer from one ground state to the other (SPRINT)\cite{Rosenblum2017,Pinotsi2008}. After the each scattering event, the atom can only scatter a photon that is coming from the opposite side and will not couple any more to a photon coming from the same direction. \\
For optimal SPRINT performance, the intrinsic loss of the resonator must be considerably smaller than $\kappa_\text{ext}$. Furthermore, the absolute values of the coupling strengths of the two transitions must be equal. Finally, a significant coupling strength is required, to ensure that the spontaneous emission is primarily directed into the resonator, rather than into free space.

The SPRINT mechanism has several interesting applications:
If the atom is in the $\ket{m_F=-1}$ ground state, light coming from the right will be transmitted. However, a single photon coming from the left, will transfer the atom into the $\ket{m_F=+1}$ ground state. In this process the photon will be reflected. From now on the atom will be transparent for all further photons coming from the left.
The system therefore behaves as a symmetric single-photon activated switch capable of routing a photon from any of its two inputs to any of its two outputs. In Ref.~\cite{Shomroni2014}, a single reflected control photon changed the switch from high reflection (R$\sim$65\%) to high transmission (T$\sim$90\%).\\
SPRINT can be employed for single photon extraction. When the atom is initialized in the ground state $\ket{m_F=-1}$, the first photon of an incoming $n$-photon pulse impinging from the left will be reflected, and the atom will be transferred into the other ground state $\ket{m_F=+1}$. From now on the atom will be transparent for all remaining photons of the pulse. Thus, $n-1$ photons of the pulse will be transmitted \cite{Rosenblum2016}.
\\
Furthermore, this scheme is the base for a deterministically operating heralded quantum memory  \cite{Lin2009} and a passive, universal photon--photon quantum gate \cite{Koshino2010}. In Ref.~\cite{Bechler2018} a passive photon-atom qubit swap-operation was demonstrated with $\sim$70\% fidelity and efficiency.

\bibliography{bibtex}

\end{document}